\documentclass[twocolumn]{aastex631}

\usepackage{graphicx}	
\usepackage{amsmath}	
\usepackage{booktabs}

\definecolor{salmon}{RGB}{239, 99, 81}
\definecolor{turchese}{RGB}{7, 190, 184}
\definecolor{smeraldo}{RGB}{85, 166, 48}
\definecolor{aqua}{RGB}{69.105, 182.070, 156.060}
\definecolor{redss}{RGB}{214, 40, 40}
\definecolor{purp}{rgb}{0.5, 0.0, 0.5}

\def\gaia{\textit{Gaia}}

\def\mudist{$\mu_{\rm dist}$}

\begin{document}

\title{Parameter Estimation for Open Clusters using an Artificial Neural Network with a QuadTree-based Feature Extractor}

\correspondingauthor{Giovanni Carraro}
\email{giovanni.carraro@aas.org}

\author[0000-0001-5831-1889]{Lorenzo Cavallo}
\affiliation{Dipartimento di Fisica e Astronomia, Università di Padova, Vicolo dell’Osservatorio 3, 35122 Padova, Italy}
\affiliation{INAF – Padova Observatory, Vicolo dell’Osservatorio 5, 35122 Padova, Italy}

\author[0000-0002-9760-6249]{Lorenzo Spina}
\affiliation{Dipartimento di Fisica e Astronomia, Università di Padova, Vicolo dell’Osservatorio 3, 35122 Padova, Italy}
\affiliation{INAF – Padova Observatory, Vicolo dell’Osservatorio 5, 35122 Padova, Italy}
\affiliation{INAF, Osservatorio Astrofisico di Arcetri, Largo E. Fermi 5, I-50125 Firenze, Italy}

\author[0000-0002-0155-9434]{Giovanni Carraro}
\affiliation{Dipartimento di Fisica e Astronomia, Università di Padova, Vicolo dell’Osservatorio 3, 35122 Padova, Italy}

\author[0000-0003-4486-6802]{Laura Magrini}
\affiliation{INAF, Osservatorio Astrofisico di Arcetri, Largo E. Fermi 5, I-50125 Firenze, Italy}

\author[0000-0003-3793-8505]{Eloisa Poggio}
\affiliation{Université Côte d’Azur, Observatoire de la Côte d’Azur, CNRS, Laboratoire Lagrange, Bd de l’Observatoire, CS 34229, 06304 Nice Cedex 4, France}
\affiliation{INAF – Osservatorio Astrofisico di Torino, Via Osservatorio 20, 10025 Pino Torinese, TO, Italy}

\author[0000-0001-8726-2588]{Tristan Cantat-Gaudin}
\affiliation{Institut de Ciències del Cosmos (ICCUB), Universitat de Barcelona (IEEC-UB), Martí i Franquès 1, 08028 Barcelona, Spain}
\affiliation{Max Planck Institute for Astronomy, Königstuhl 17, 69117 Heidelberg, Germany}

\author[0000-0003-3784-5245]{Mario Pasquato}
\affiliation{Dipartimento di Fisica e Astronomia, Università di Padova, Vicolo dell’Osservatorio 3, 35122 Padova, Italy}
\affiliation{Département de Physique, Université de Montréal, Montreal, Quebec H3T 1J4, Canada}
\affiliation{Mila - Quebec Artificial Intelligence Institute, Montreal, Quebec, Canada}
\affiliation{Ciela, Computation and Astrophysical Data Analysis Institute, Montreal, Quebec, Canada}

\author[0000-0001-8808-0073]{Sara Lucatello}
\affiliation{Institute for Advanced Studies, Technische Universität München, Lichtenbergstraße 2 a, 85748, Garching bei München, Germany}

\author[0000-0001-7939-5348]{Sergio Ortolani}
\affiliation{Dipartimento di Fisica e Astronomia, Università di Padova, Vicolo dell’Osservatorio 3, 35122 Padova, Italy}
\affiliation{INAF – Padova Observatory, Vicolo dell’Osservatorio 5, 35122 Padova, Italy}

\author[0000-0002-2179-9363]{Jose Schiappacasse-Ulloa}
\affiliation{Dipartimento di Fisica e Astronomia, Università di Padova, Vicolo dell’Osservatorio 3, 35122 Padova, Italy}
\affiliation{INAF – Padova Observatory, Vicolo dell’Osservatorio 5, 35122 Padova, Italy}




\begin{abstract}
With the unprecedented increase of known star clusters, quick and modern tools are needed for their analysis. In this work, we develop an artificial neural network trained on synthetic clusters to estimate the age, metallicity, extinction, and distance of \gaia\ open clusters. We implement a novel technique to extract features from the colour-magnitude diagram of clusters by means of the QuadTree tool and we adopt a multi-band approach. We obtain reliable parameters for $\sim 5400$ clusters\footnote{The catalogue of clusters analysed in this work is available at \url{https://phisicslollo0.github.io/cavallo23.html}. See Appendix \ref{sec:appendix} for more details.}. We demonstrate the effectiveness of our methodology in accurately determining crucial parameters of \gaia\ open clusters by performing a comprehensive scientific validation. In particular, with our analysis we have been able to reproduce the Galactic metallicity gradient as it is observed by high-resolution spectroscopic surveys. This demonstrates that our method reliably extracts information on metallicity from colour-magnitude diagrams (CMDs) of stellar clusters. For the sample of clusters studied, we find an intriguing systematic older age compared to previous analyses present in the literature. This work introduces a novel approach to feature extraction using a QuadTree algorithm, effectively tracing sequences in CMDs despite photometric errors and outliers. The adoption of ANNs, rather than Convolutional Neural Networks, maintains the full positional information and improves performance, while also demonstrating the potential for deriving clusters' parameters from simultaneous analysis of multiple photometric bands, beneficial for upcoming telescopes like the Vera Rubin Observatory. The implementation of ANN tools with robust isochrone fit techniques could provide further improvements in the quest for open clusters' parameters. 
\end{abstract}

\keywords{Open star clusters (1160) -- Gaia (2360) -- Neural networks (1933)}


\defcitealias{CG20}{CG20}
\defcitealias{D21}{D21}
\defcitealias{H23}{HR23}

\section{Introduction} \label{sec:intro}
Open stellar clusters are groups of stars formed together from the same molecular clouds. They are simple stellar populations whose members are born approximately at the same time, share the same location in the sky, distances, kinematics and the same initial chemical composition \citep[e.g.][]{DeSilva06,DeSilva07,BlandHawthorn10,Armillotta18}. Given all these properties, open clusters have long been considered benchmarks in the context of the determination of stellar properties such as ages, distances and chemical composition. In fact, these quantities can be derived for cluster members with extremely high precision compared to what is normally achievable for individual field stars.

For that reason, open clusters are the ideal laboratories for a large variety of studies concerning stellar evolution and the Galaxy as a whole \citep{Adams01,Zwart01,Krumholz19,Wright20,CantatGaudin22}.
Numerous stellar evolution processes can be investigated with open cluster members: star formation \citep{Lada03}, progression in Hertzsprung-Russell diagram \citep{Vandenberg83,Babusiaux18}, evolution of rotational properties \citep{Meibom11,Gruner20,GodoyRivera21}, mixing episodes \citep{Magrini21a,Dumont21}, and diffusion processes \citep{Souto18,BertelliMotta18}. Clusters have also long been used as tracers of the structure and evolution of the Milky Way \citep{Moffat73,Janes82,Dias05,Moitinho06,Carraro09,Carraro10,Xu18,Reid19,CG20,Poggio21}, the distribution of element across the Galactic disk \citep{Janes79,Friel02,Donor18,Donor20} and offered perspectives on how this distribution has evolved with time \citep{Carraro98,Myers22,Spina22a, Magrini2023A&A...669A.119M}. Thanks to their ages and precisely determined chemical patterns, open clusters are excellent calibrators for chemical clocks: powerful, alternative, chemically-based diagnostics of ages for field stars, such as the carbon-to-nitrogen abundance ratio \citep{Casali19,Spoo22}. 

The entire life cycle of open clusters is deeply connected to several aspects of Galaxy evolution. For instance, their birth and death can be strongly influenced by large-scale non-axisymmetric features in the Milky Way, such as spiral arms, as well as giant molecular clouds on a small scale. \citep{Piskunov06,Morales13,Piskunov18,Wright20,Anders21}.
Therefore, it has been proven to be important to understand the dynamical evolution and dissipation of open clusters \citep{Bravi18,Yeh19,Carrera19,Tang19,Meingast21,Pang21,Casamiquela22a}, and also comparing their demography to that of field stars \citep{Soubiran18,Spina20}. 

Open clusters are also critical in testing the possibility of using the chemical composition of stars to identify the environment where they formed \citep{Freeman02,Mitschang13,BlancoCuaresma15,GarciaDias19,Spina22b}. Beyond their value in understanding the Galaxy and stellar evolution, open clusters are frequently used as reference objects to assess the quality of the large astronomical datasets \citep{Babusiaux22,RecioBlanco22,Randich22} and calibrators for stellar yields models \citep{Maiorca12,Magrini21b}. 

All this broad variety of studies is grounded on reliable determinations of open cluster properties such as distance, age, and extinction.
The advent of Gaia astrometric and photometric data \citep{Prusti16} resulted in
substantial advancement in our knowledge of the open cluster population. Current astrometric data allow a much more complete census of stellar associations and of their stellar members than pre-$Gaia$ catalogues \citep[e.g.][]{Dias02,Kharchenko13}. All that, in conjunction with the precise Gaia photometry, allows astronomers to appreciate tight and clear sequences in colour-magnitude diagrams, outlined by the members of such stellar populations, fundamental pieces of information from which we can extract key physical parameters of open clusters. As a consequence of this progress, automated tools for the characterisation of open clusters based on advanced statistical methods have become increasingly common \citep[e.g.][]{CG20,D21,H23}.\\
In June 2022 \gaia\ \textit{Collaboration} has published the third data release \citep[DR3,][]{gaiadr3} that contains the photometry ($G$, $G_{\rm BP}$, and $G_{\rm RP}$) of more then 1.5 billion objects. From this data, clustering algorithms are discovering more and more star clusters \citep[e.g.][]{H23,Perren2023} for which a proper parameter estimation is required. To do that, one method often used is the so-called isochrone fitting, which provides good results, but is usually performed case by case and so is not efficient when used in samples that contain hundreds or even thousands of clusters. 

Recently, \citet{D21} \citepalias[hereafter][]{D21} have determined the parameters, via isochrone fitting procedure, of 1743 open clusters contained in Gaia DR2. This kind of procedure also incurs significant computational overhead. To address this challenge and overcome the high computational cost, \citetalias{D21} make use of strong priors in clusters' extinction and metallicity, which can guide the procedure towards more efficient and focused computations, resulting in improved performance. However, the introduction of such priors can introduce biases into the analysis. Furthermore, the entire procedure is susceptible to the presence of faint and red non-member stars, which can significantly impact its performance and reliability.\\
In recent years, due to the ever-growing number of open clusters and the considerable computational demands of the isochrone fitting procedure, several researchers have explored the utilisation of machine learning algorithms to estimate the parameters of open clusters. \citet{CG20} \citepalias[hereafter][]{CG20} have used an Artificial Neural Network (ANN) \citep{ANNcite} to estimate age, extinction, and distance of 1867 \gaia\ open clusters. Their neural network has been trained using well-known clusters that possess precisely defined parameters. Considering the utilisation of an observational sample of open clusters which are typically younger than a few 100 Myr, there may exist potential biases in their analysis. \\
\citet{H23} \citepalias[hereafter][]{H23} employed the Gaia DR3 dataset to conduct a comprehensive search for open clusters across the entire sky. They utilised the Hierarchical Density-Based Spatial Clustering of Applications with Noise (HDBSCAN) algorithm \citep{HDBSCAN} to identify potential cluster candidates. These candidates were then subjected to validation using a quasi-Bayesian Convolutional Neural Network (CNN) designed to distinguish genuine clusters from spurious ones. In fact, variable extinction - which is a frequent condition near the Galactic midplane - and random density fluctuations can make unbound groups of stars appear as relative overdensities in sky coordinates, parallaxes and proper motions \citep[e.g.][]{Kounkel20}. Furthermore, a similar neural network was trained to estimate the ages, extinctions, and distances of the identified clusters. By employing these techniques, \citetalias{H23} generated a unified catalogue of star clusters consisting of 7167 entries. Their training set was assembled with synthetic clusters, which suggests a potential absence of biases compared to the case of \citetalias{CG20}.\\
In this work, we employ a group of ANNs trained on synthetic OCs to estimate the age, metallicity, extinction, and distance modulus of a sample of clusters obtained from \citetalias{H23}. Our method of analysis has several differences from the techniques used in the past. The most significant element of novelty is that we use multiple colour-magnitude diagrams (CMDs) constructed using \gaia\ and 2MASS photometry. By incorporating additional information, our approach provides a more comprehensive analysis of the clusters' properties. Also, developing and testing an algorithm that conjointly analyses multiple photometric bands is strategic in view of next-generation telescopes, such as the Vera Rubin Observatory which will provide photometry from six filters for $\sim 20\times$10$^9$ stars \citep{LSST17}. 

Instead of directly feeding the neural networks with the CMD images, with our approach, we first extract the relevant features of the distributions of stars within these diagrams with the QuadTree algorithm \citep{Schiappacasse2023arXiv230615487S} and then we feed the neural-net with this information only. By doing so we significantly reduce the complexity of the network that will be trained compared to typical networks that need to process full images. Besides all that, our approach is different in many other ways from past techniques of analysis. It is by exploring and developing new techniques that we enhance the accuracy and comprehensiveness of parameter estimation for the population of Galactic open stellar clusters.\\
This paper is structured as follows. In Section~\ref{sec:Data} we describe the dataset employed in our study. In Section~\ref{sec:Artificial_neural_network} we give details on the model we use for the parameter estimation and how it has been trained. The results and comparison with previous studies are discussed in Section~\ref{sec:Comparison_literature}. In Section~\ref{sec:Scientific_validation} we present a scientific validation of our methodology to derive clusters' parameters. Finally, conclusions are discussed in Section~\ref{sec:Conclusions}.

\section{DATA}\label{sec:Data}

Recently, \citetalias{H23} presented a homogeneous sample of open clusters (see Section~\ref{sec:Catalogue_of_OC}), providing estimates for age, extinction, and distance. Building upon their work, we extend their analysis by employing a distinct method to estimate age, extinction, distance, and metallicity for the same sample of clusters. To accomplish this, we utilise a group of ANNs trained with a set of synthetic open clusters generated using a numerical Monte Carlo code developed specifically for this purpose (see Section~\ref{sec:Synthetic_CMDs}). 

In this Section, we describe the catalogue of open clusters for which we estimate the parameters and the training set we used for this purpose. 
\begin{figure}
    \centering
    \includegraphics[width=0.5\textwidth]{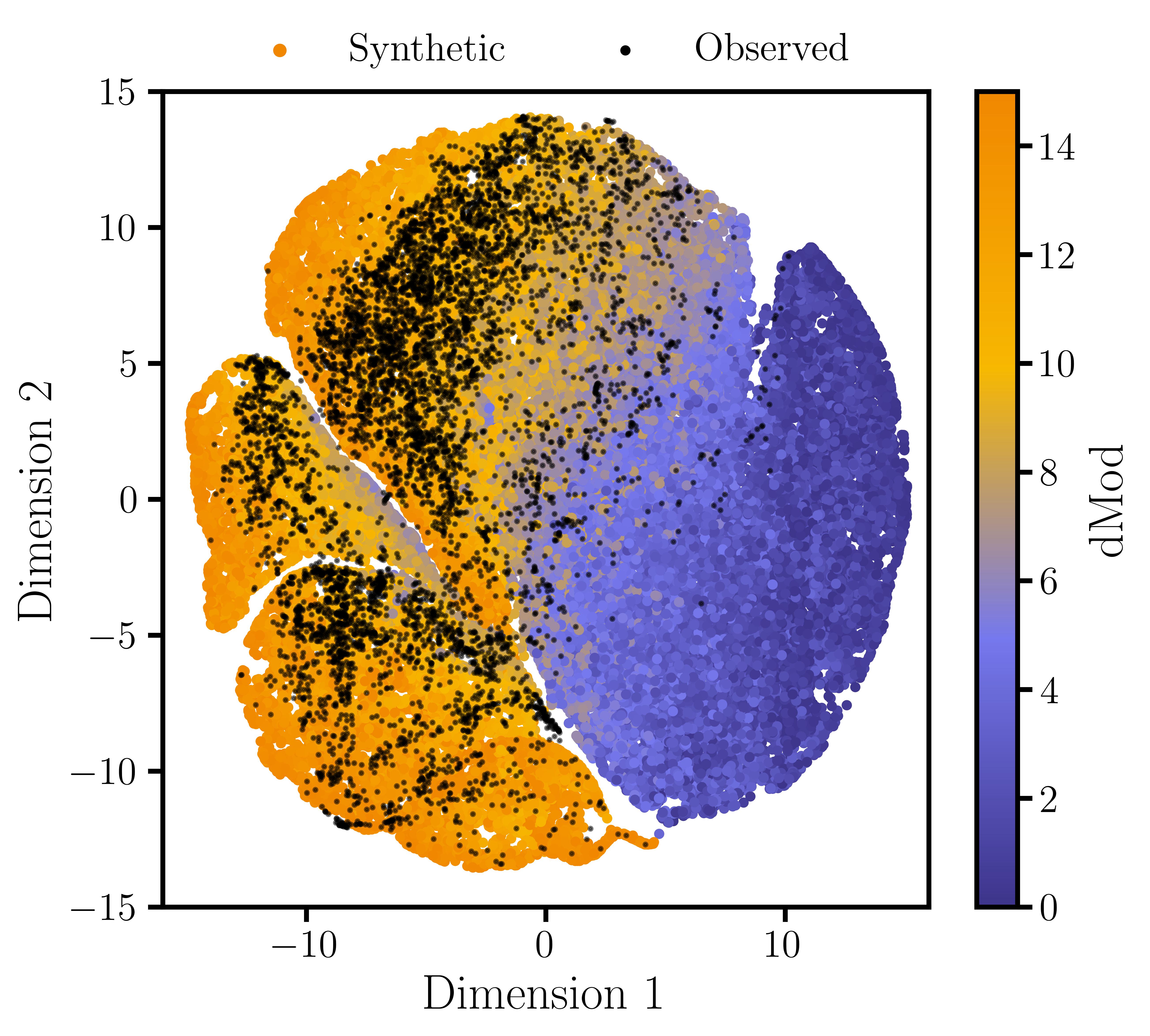}
    \caption{Superimposed t-SNE maps of synthetic and observed OCs. Synthetic clusters are colour-coded based on their distance modulus, while observed clusters are plotted with black dots.}
    \label{fig:tSNE}
\end{figure}

\subsection{Catalogue of Open Clusters} \label{sec:Catalogue_of_OC}
\citetalias{H23} have conducted an all-sky blind search for open clusters in the \gaia\ DR3 dataset using the Hierarchical Density-Based Spatial Clustering of Applications with Noise (HDBSCAN) algorithm. Candidates identified by HDBSCAN have been validated with a quasi-Bayesian convolutional neural network (CNN) trained to discriminate between real and fake clusters from pixelled images of their CMDs. A similar neural network is then used to infer the age, extinction, and distance of detected clusters. They then produced a homogeneous star cluster catalogue composed of 7167 clusters ($\sim 2400$ of them were not present in the previous catalogues). From the initial catalogue, we select only star members with $G\leq18$. It is important to note that the full catalogue contains open clusters, globular clusters, and moving groups and that, as it results from \citetalias{H23}'s analysis, some of these 7167 stellar associations may not be real.\\
From \citetalias{H23} we select the stellar associations with $\geq10$ members ending up with 6413 clusters. In addition to the Gaia photometry, in our study, we also use the 2MASS $J$ and $H$ magnitudes for each member of these clusters, when available. More specifically, we use only the 2MASS magnitudes that have signal-to-noise ratio greater than $7$ (2MASS photometric quality flag equal to A or B) and photometric errors $\leq 0.3$ mag.

\subsection{Synthetic colour-magnitude diagrams} \label{sec:Synthetic_CMDs}
An isochrone is the locus of points in the CMD of stars sharing the same age and chemical composition. The isochrone alone is not informative on the number of stars that are in a certain evolutionary stage (i.e. that populate different zones of the CMD). This information is given by the initial mass function (IMF) and the initial mass of the cluster. Moreover, in order to develop a synthetic colour-magnitude diagram (CMDs) of an open cluster the effects on the CMD of extinction, distance, and photometric errors have to be considered.\\
In this work, we develop a numerical Monte Carlo code which simulates a CMD of an open cluster (i.e. simple stellar population) of age $t$, metallicity [Fe/H], affected by an extinction $A_V$, situated at a distance $d$, and composed by a number of members $N_{\rm star}$. For each star in our synthetic clusters, we compute the apparent magnitude in the following bands: $G$, $G_{\rm BP}$, $G_{\rm RP}$, $J$, and $H$.\\
In the first step, the code searches among a grid of theoretical isochrones the one of a stellar population of age $t$ and metallicity [Fe/H]. For the theoretical grid of models, we use version 1.2S of PARSEC isochrones \citep{Bressan2012MNRAS.427..127B, Chen2014MNRAS.444.2525C, Tang2014MNRAS.445.4287T, Chen2015MNRAS.452.1068C}. The selected isochrone gives the relation between the absolute magnitude $M_{\lambda}$ (where $\lambda$ indicates a generic band) and the mass of the star. At this point, we transform the absolute magnitudes of the isochrone into apparent ones ($m_{\lambda}$) as:
\begin{equation}
    m_{\lambda} = M_{\lambda} + {\rm dMod} + A_V \times {\rm c}_{\lambda}
\end{equation}
where $M_{\lambda}$ is the absolute magnitude in the generic band $\lambda$, dMod is the distance modulus (dMod$=5\log(d)-5$), $A_V$ is the extinction in the $V$ band, and $c_{\lambda}$ is the ratio of the extinction coefficients in the $\lambda$ and $V$ band. The coefficients $c_{\lambda}$ depend both on $A_V$ of the cluster and the effective temperature ($T_{\rm eff}$) of stars and thus it needs to be computed star-to-star. To do that, we used the python package \texttt{pysvo}\footnote{\url{https://github.com/fjaellet/pysvo}} \citep{SVO_1_2012ivoa.rept.1015R} that allows you to handle data coming from the Theoretical Model Services maintained by the Spanish Virtual Observatory. For this computation, we assume an extinction curve with $R_V=3.1$ \citep{Cardelli1989ApJ...345..245C}.\\
Then, we generate $N_{\rm stars}$ synthetic stars with a mass randomly chosen accordingly with the \citet{Salpeter1955ApJ...121..161S} IMF. With the isochrone computed in the first step, we obtain the apparent magnitude of each star.\\
To train our neural network we use a sample that consists of 50~000 synthetic open clusters generated with the CMD generator described above. The synthetic population of OCs has been generated with parameters distributed as reported in Table~\ref{tab:synthetic_pop_params}. For each synthetic OC, we also derive its expected parallax as the inverse of the distance.\\\\
During the development of our tool, we considered well-known astronomical phenomena that have the potential to sculpt the CMD and, consequently, influence our ANN results. Phenomena such as stellar rotation, binaries \citep[see][]{Cordoni2023A&A...672A..29C, Donada2023A&A...675A..89D}, extended main-sequence turn-off, and differential extinction, among others, undeniably play important roles in shaping the CMD and can introduce complexities and potential sources of error when not accounted for.\\
Our decision to employ a simplified, yet plausible, model to test the effectiveness of the QuadTree algorithm in extracting CMD features, and the overall capability of the ANN to efficiently extract parameters like logAge, Av, dMod, and notably, the metallicity, was driven by a strategic choice to establish a robust baseline.\\
It is important to acknowledge that the aforementioned astronomical processes have a tangible impact on the CMD and omitting them could indeed be a source of error. However, it is crucial to highlight a pivotal aspect of our ANN: it does not “see” the position of individual cluster members. Instead, it processes a series of coordinates that delineate zones of point over-densities, thereby lacking the capability to discriminate between a CMD that presents a sequence of binary stars and one where the main sequence is more sparse than predicted by theoretical models.\\

\subsection{t-Distributed Stochastic Neighbour Embedding}
Before we use our synthetic sample of OCs to train the ANN, we need to check whether or not the training set is a good generalisation of all the possible observations. As we anticipated in Section~\ref{sec:intro}, the ANN that we use to determine the OCs' parameters receives features extracted from clusters' CMDs. More specifically, it is fed with 63 input features that are extracted from the set of three CMDs built for each of the clusters on the basis of Gaia and 2MASS photometry (see Section \ref{sec:Architecture} for details), plus an additional input feature which is the parallax. The procedure of feature extraction from the CMDs is described in Section~\ref{sec:QuadTrees}. Here, we aim to demonstrate that distributions of all the features we extract from the observed CMDs overlap with those of the same features extracted from the synthetic CMDs.\\
In the fields of data analysis and machine learning, understanding and visualising complex high-dimensional data sets is a critical task. Traditional visualisation techniques often struggle to effectively capture the intricate relationships and structures hidden within such data. However, t-Distributed Stochastic Neighbour Embedding (t-SNE) \citep{vanderMaaten2008} maps offer a powerful approach to address this challenge.\\
t-SNE maps provide a means of visualising high-dimensional data in a reduced dimensional space while attempting to preserve the inherent relationships and structures among the data points, especially at small scales. This is accomplished by minimising a measure of dissimilarity between the distance distribution in the high-dimensional space and in the low-dimensional projection; an accessible explanation of the algorithm is provided by \cite{wattenberg2016how}.
The resulting t-SNE map provides a visually intuitive representation of the data, where clusters, patterns, and similarities are discernible. Points that are similar in the original space tend to be grouped together in the t-SNE map, enabling analysts to identify distinct clusters or regions of interest.\\
In Fig.~\ref{fig:tSNE} we plot the superimposed t-SNE maps obtained by the \gaia\ CMDs of synthetic (colour-coded accordingly their distance modulus) and observed (black dots) clusters. In that plot, we notice that the clusters of \citetalias{H23} are perfectly embedded in the synthetic sample generated by us and that we use to train the neural network. It is also clear that the observed clusters cover only a portion of the area filled by the synthetic clusters. 
Since observing an individual t-SNE embedding does not per se guarantee that these relations are also reflected in our original feature space, we have examined various plots similar to Fig.~\ref{fig:tSNE} obtained for different values of the perplexity hyperparameter, which sets the effective scale at which structure is rendered faithfully. The results appear robust to such changes.

\section{Artificial neural network} \label{sec:Artificial_neural_network}
An ANN is an algorithm that maps input data into output parameters through a network of nodes and weights. In our case, we feed the ANN with parallaxes of OCs and information extracted from their CMDs (input) and obtain their age, metallicity, extinction, and distance modulus (output). Between the input and output layers, hidden layers of neurons are placed. These layers increase the complexity of the network and its predictive potential. However, as the network becomes more complex, it develops a susceptibility to over-fitting. In fully connected ANNs (as the one developed in this work), each neuron in a layer receives from (all) the nodes of the previous layer an input, that is transformed with a non-linear function and transmitted to the neurons of the next layer.\\
In this work, we train an ANN with synthetic OCs and then use it to estimate the parameters of the sample of clusters from \citetalias{H23}. Below we describe the method used to build the sample of synthetic OCs used to train our ANN (Section~\ref{sec:Synthetic_CMDs}), its input layer (Section~\ref{sec:QuadTrees}), architecture (Section~\ref{sec:Architecture}), training and performance (Section~\ref{sec:training_performance}).

\begin{figure}
    \centering
    \includegraphics[width=0.5\textwidth]{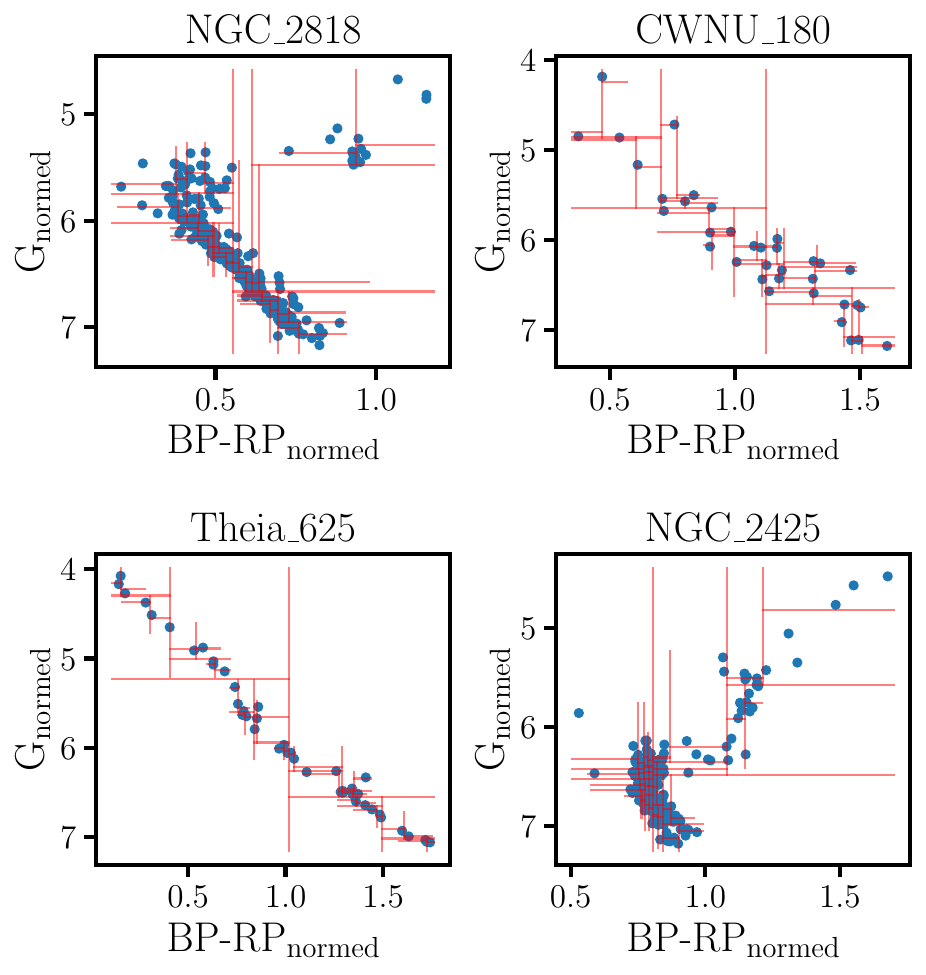}
    \caption{QuadTrees of Gaia-CMD of four clusters contained in \citet{H23}. Note that magnitudes and colours are standardised as described in Section~\ref{sec:QuadTrees}}.
    \label{fig:QuadTree}
\end{figure}
\begin{figure}
    \centering
    \includegraphics[width=0.5\textwidth]{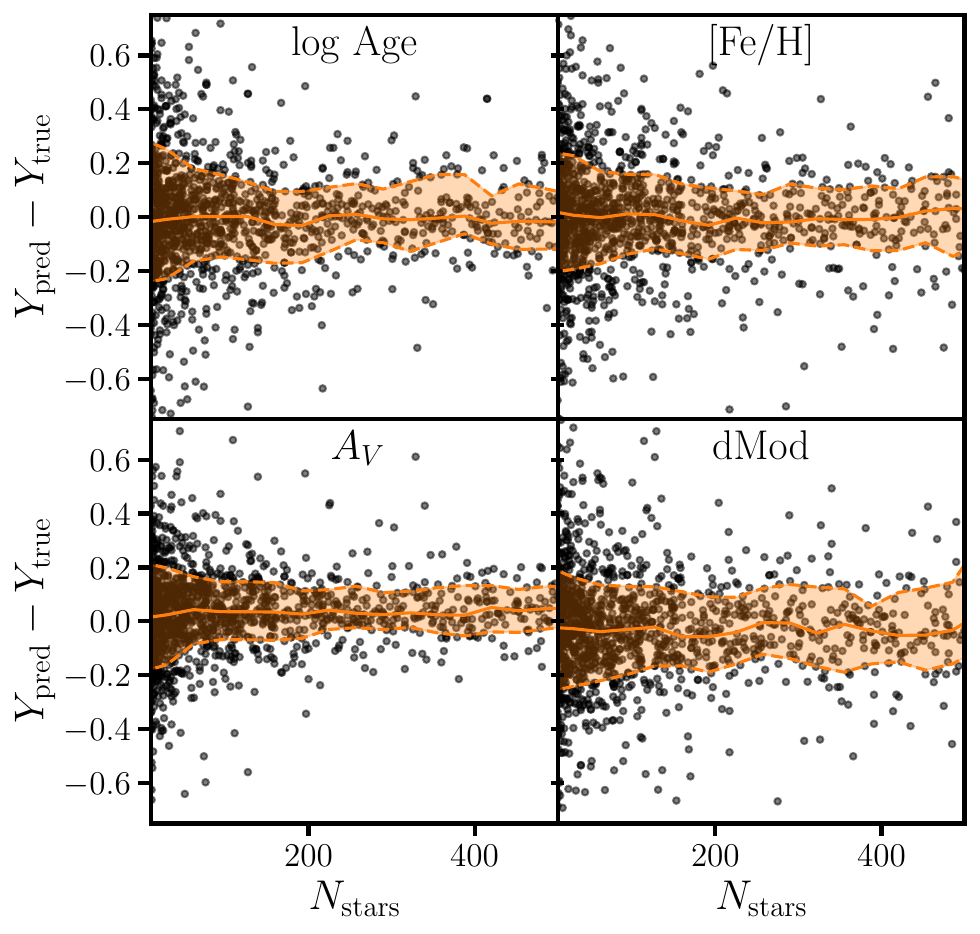}
    \caption{Difference between the parameters estimated by our ANN and the true ones as a function $N_{\rm stars}$ for $\sim 2000$ synthetic clusters contained in the test sample (black dots). In orange, we plot the mean and standard deviation of $Y_{\rm pred}-Y_{\rm true}$.}
    \label{fig:performance}
\end{figure}

\begin{table}
	\centering
	\caption{Distribution of parameters of our synthetic clusters used to train our ANN.}
	\label{tab:synthetic_pop_params}
	\begin{tabular}{cc} 
		\hline
		\textbf{Parameter} & \textbf{Distribution}\\
        \hline
		  logAge & $\mathcal{U}(6.7,10.1)$ \\
        $[$Fe/H$]$ & $\mathcal{U}(-0.8,0.6)$ \\
        $A_{\rm V}$ & $\mathcal{U}(0,5)$ \\
        dMod & $\mathcal{U}(0,15)$\\
        log $N_{\rm stars}$ & $\mathcal{U}(1.0,3.5)$\\
		\hline
	\end{tabular}
\end{table}

\begin{figure*}
    \centering
    \includegraphics[width=0.9\textwidth]{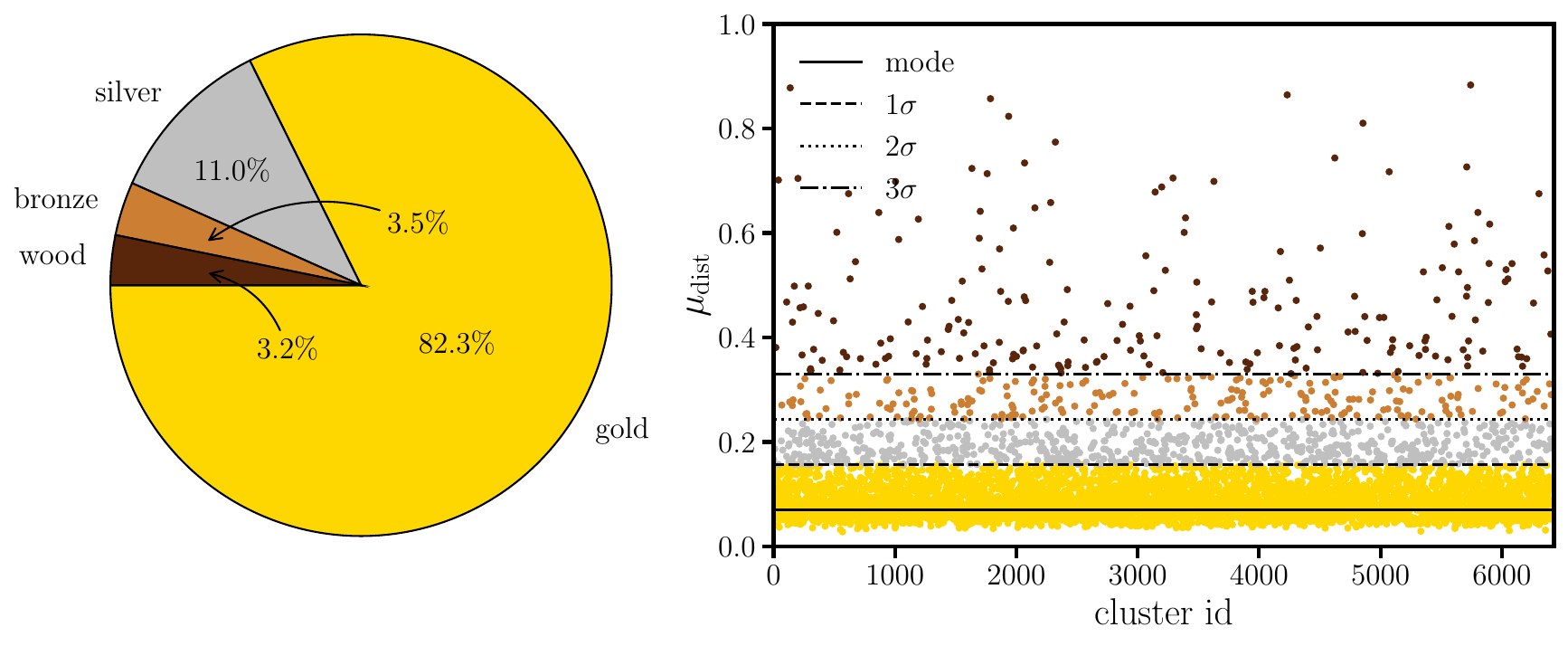}
    \caption{Left: fraction of clusters contained in gold ($82.3\%$), silver ($11.0\%$), bronze ($3.5\%$), and wood ($3.2\%$) sample. Right: distribution of \mudist\ for the 6413 clusters analysed in this work. Clusters are coloured according to their 'quality' sample (gold, silver, bronze, or wood).}
    \label{fig:g_s_b_w_sample}
\end{figure*}

\begin{figure}
    \centering
    \includegraphics[width=0.45\textwidth]{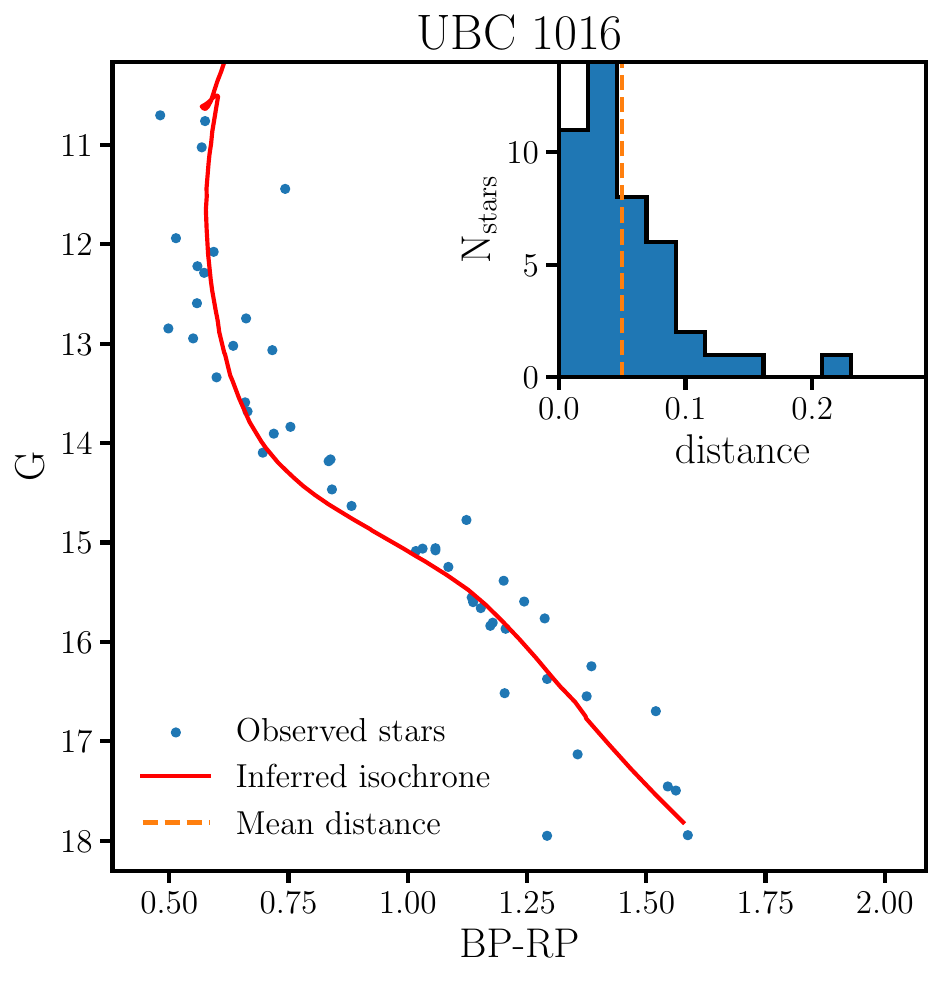}
    \caption{Example of the computation of the quality metric developed in this work. With blue dots, the members of the open cluster UBC~1016 on which we superimpose the isochrone inferred by the ANN (plotted with a solid red line). In the top right corner, we plot the histogram of the distances between the inferred isochrone and the members of the cluster. On the histogram, we plot with a vertical dotted orange line the mean value of the distribution of distances. }
    \label{fig:quality_metric}
\end{figure}

\subsection{Features extraction} \label{sec:QuadTrees}
To build the input layer of our ANN we need to pre-process the CMD of each cluster and transform it into a one-dimensional array. To do that we compute the QuadTree of CMDs.\\
A QuadTree is a tree data structure which is used to represent object distribution in 2-dimensional space. It takes a distribution of points on a xy-plane, in the first level of the tree the algorithm divides the xy-plane into four different areas that contain the same number of points. It thus saves the coordinates of the edges of these areas and so it obtains one x-coordinate ($x_1$) and two y-coordinates ($y_1,y_2$). In the second level, it repeats the procedure in each of the four areas of the first level. It then obtains again one x-coordinate and two y-coordinates per each area. At the end of the second level, it obtains two arrays of coordinates: ($x_1^{1st}, x_2^{2nd},\dotsc,x_5^{2nd}$) and ($y_1^{1st}, y_2^{1st}, y_3^{2nd},\dotsc,y_{10}^{2nd}$ ); where $1st$ and $2nd$ label coordinates that come from first and second levels, respectively. It repeats the same procedure also for the third and last level and it obtains two arrays:  ($x_1^{1st},\dotsc,x_{21}^{3rd}$) and ($y_1^{1st},\dotsc,y_{42}^{3rd}$ ). Then we concatenate them into a unique array with 63 entries.\\
In a typical CMD, the absolute star-to-star difference in magnitude is larger than the one in colour. In this situation, the ANN may weigh more the differences in magnitude and less the one in colour. To avoid this situation we standardise the CMD diagrams by dividing each quantity by its standard deviation computed on the whole sample of synthetic CMDs. We then compute the QuadTrees of these standardise CMDs.\\
In Fig.~\ref{fig:QuadTree} we plot the QuadTrees of four selected clusters. It is possible to notice that the algorithm is able to trace the sequence of OCs with a fair number of stars, as well as in the case of OCs with few tens of members. 
\subsection{The neural network architecture}\label{sec:Architecture}
In this work, we tested many ANNs with various architectures, input layers and characteristics. In the following lines, we are going to describe the combination of characteristics that leads to the best performance.\\
Our ANN is made of fully-connected layers and is implemented in TensorFlow \citep{Abadi2016arXiv160304467A}. The input layer receives the $63\times 3$ features produced by the QuadTree (see Section~\ref{sec:QuadTrees}) extracted from the three different CMDs ($(G_{\rm BP} - G_{\rm RP})$ vs $G$, ($(G_{\rm BP} - G_{\rm RP})$) vs $J$, and ($J-H$) vs $J$) of the cluster that we want to study plus the mean parallax of its members. In particular, the input layer has 190 neurons dived in $3\times63$ entries from the flattened QuadTrees of the three CMDs and one with the mean parallax of the cluster. As described in Section~\ref{sec:QuadTrees}, the QuadTrees are computed on standardised CMDs. In this layer, we add a Dropout function that randomly switches off the $10\%$ of neurons of the layer. We verified that this helps the ANN to deal with contaminants present in CMDs of real OCs that produce out-layers in the QuadTree.\\
After the input layer, we place five hidden layers, each one composed of 500 neurons. For the activation function of these hidden layers, we use the Leaky version of the Rectified Linear Unit (LeakyReLU) with $\alpha=0.3$. As output, the ANN gives the logarithm of the age ($\log t$), metallicity ([Fe/H]), visual extinction ($A_{\rm V}$), and distance modulus (dMod) of a cluster given its CMDs and the mean parallax.\\

\subsection{Training and performance}\label{sec:training_performance}
In general, an ANN is deterministic, i.e. at a certain input will produce a certain output without a metric that represents how well determined that output is. In order to estimate that metric we trained, 300 ANNs on different sub-samples of 10 000 clusters randomly selected among the total synthetic sample (50 000).\\ 
To train each ANN we divide the 10~000 synthetic clusters in train ($\sim 80\%$) and test ($\sim 20\%$) samples. The train is performed with the Adam solver \citep{Adam} that optimises the mean squared error (MSE) loss function.\\
In order to evaluate the reliability of the ANN to obtain the parameter of a cluster given its CMDs and parallax we analyse its performance on $\sim 2000$ synthetic clusters that are part of the test sample. In Fig.~\ref{fig:performance} we plot the difference between the parameters estimated by our ANN and the real ones. As expected we notice that the parameters of clusters with fewer members are more uncertain. 
On average, our ANN estimate logAge, [Fe/H], $A_{\rm V}$, and distance modulus with an uncertainty $\lesssim 0.2$. These uncertainties are a combination of the errors made by the ANN itself and the ambiguities generated by photometric errors (embedded in the CMDs of our synthetic clusters, see Section~\ref{sec:Synthetic_CMDs}).
By aggregating the predictions from these 300 ANNs we enhance the reliability of the estimates and gain a more comprehensive
understanding of the uncertainties introduced by our ANNs.

\begin{figure*}
    \centering
    \includegraphics[width=\textwidth]{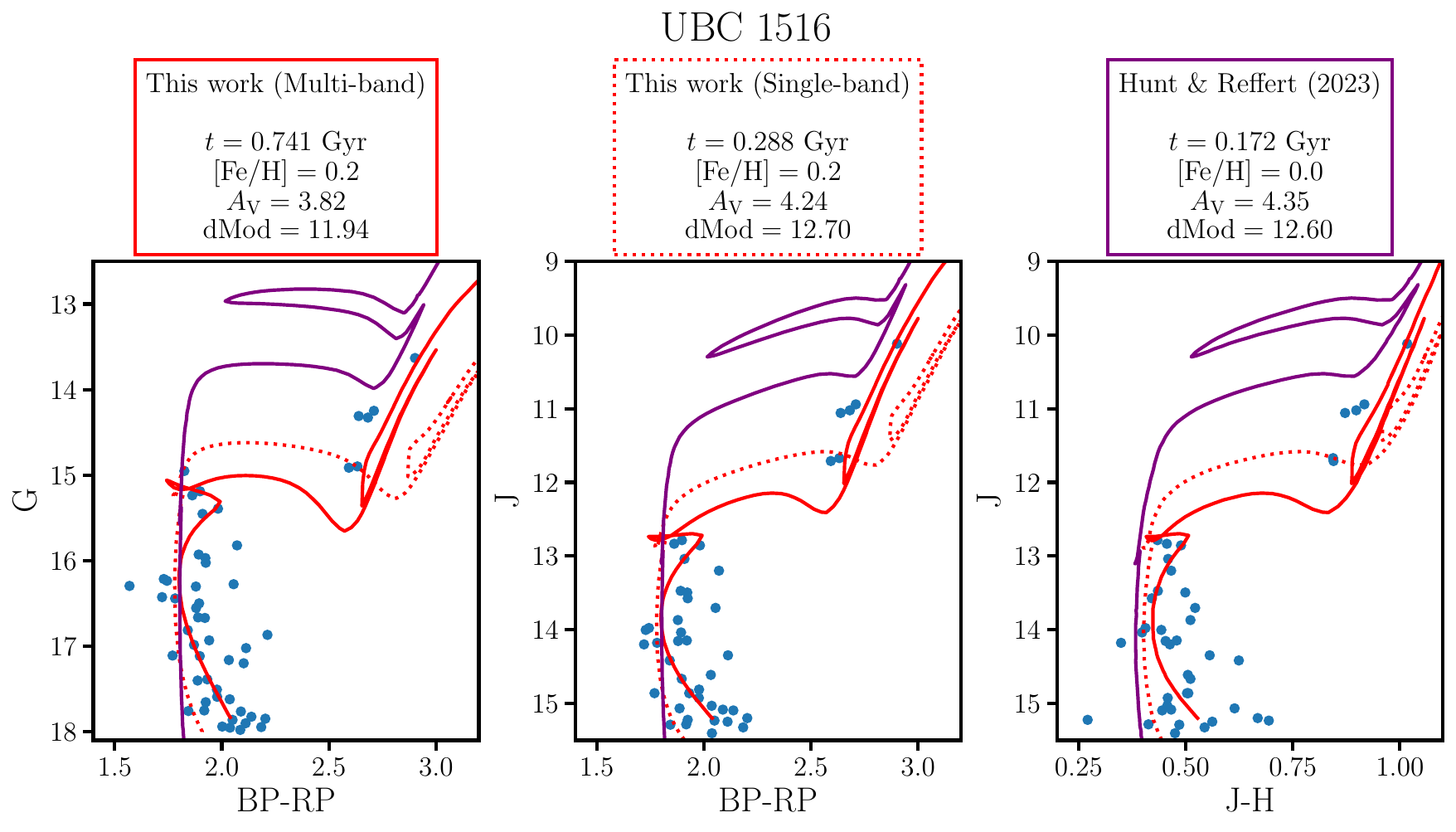}
    \caption{Isochrones and parameters for UBC~1516 inferred by \citetalias{H23} (solid purple line), our single-band ANN (dotted red line), and multi-band ANN (solid red line). The predicted isochrones are superimposed on the corresponding CMDs of the cluster. Note that the isochrone inferred by \citetalias{H23} is at solar metallicity. This value is fixed and not actually derived by their ANN.}
    \label{fig:multiband}
\end{figure*}

\subsection{Parameter estimation}
Once the 300 ANNs are trained, we proceed with the parameter estimation of the real clusters. To do so we proceed as follows: i) the three CMDs are standardised as described in Section~\ref{sec:QuadTrees}; ii) we compute the QuadTree of each CMD, merge them and add the mean parallax of cluster's members obtaining the input layer; iii) the input layer is fed to the 300 ANNs and each of them gives an estimate for logAge, [Fe/H], $A_{\rm V}$, and dMod. Thus, for each parameter, we obtain a distribution from which we extract the mean (our estimated value) and $1\sigma$ confidence interval, i.e. its uncertainty. \\
Given the large number of objects, it is impossible to perform a visual inspection of the CMDs of each cluster with theoretical isochrones computed with the estimated parameters to check if there is a good match. We thus need a metric that indicates the quality of the match between the estimated isochrone and the observed CMD. To do that, we developed a metric based on the distance of stars in the \gaia\ CMD from the isochrone predicted by our neural networks. Given the \gaia\ CMD of a cluster, for each individual cluster member, we normalise colours and magnitudes (see Section~\ref{sec:QuadTrees}) then we compute the distance of the star from the closest point of the inferred isochrone, next we average them obtaining the mean distance of the CMD from the isochrone ($\mu_{\rm dist}$). In Fig.~\ref{fig:quality_metric} we plot an example of the computation just described, for the cluster UBC~1016. \\
We divide the sample of clusters analysed in this work into four different categories based on the value of the metric (\mudist) described above: gold, silver, bronze and wood. Given the distribution of \mudist, plotted in the right panel of Fig.~\ref{fig:g_s_b_w_sample}, we compute its standard deviation $\sigma_{\mu_{\rm dist}}$ (hereafter $\sigma_{\mu_{\rm dist}}\equiv \sigma_\mu$) and subsequently we label our parameter estimate of a cluster as: i) gold, if \mudist\ is within $\sigma_{\mu}$ from the mode of the distribution; ii) silver, \mudist\ between $1\sigma$ and $2\sigma$; iii) bronze, \mudist\ between $2\sigma$ and $3\sigma$; iv) wood, \mudist\ beyond $3\sigma$. As plotted in the right panel of Fig.~\ref{fig:g_s_b_w_sample} $\sim 5300$ ($>80$\%) clusters belong to the gold sample and so their inferred isochrones have a good match with the observational data.\\
About $200$ clusters ($\sim 3\%$) belong to the wood sample, where ANN struggles to extract parameters that are in agreement with the observed CMDs. Typically the members of this wood group present CMDs that are poorly populated (low number of stars) have high star-to-star scatter in both magnitudes and colours or contain strong contaminants.

\subsection{The multi-band approach}
As discussed in the previous sections, our ANN uses multiple CMDs to estimate the parameters of \gaia\ open clusters. In particular, it uses a combination of magnitudes and colours from \gaia\ and 2MASS: $(BP-RP)$ vs $G$, $(BP-RP$ vs $J$, and ($J-H$) vs $J$. The use of these three different CMDs comes in handy to partially disentangle the degeneracies between age, metallicity, extinction, and distance taking advantage of the fact that different bands have different sensibilities on these parameters.\\
To test the improvements given by our multi-band approach, we trained another group of 300 ANNs with an input layer of 64 neurons composed by the 63-entry QuadTree array of the \gaia\ CMD ($(G_{\rm BP} - G_{\rm RP})$ vs $G$) and the parallax of the clusters. From a visual inspection of a few hundred clusters is clear that the performance of the ANNs that use only the \gaia\ CMDs is worse than in the case of the multi-band ones. In particular, the additional two CMDs seem to help the ANN to partially disentangle the degeneracy between the four parameters. In Fig.~\ref{fig:multiband} we show an insightful example of this. We plot in the three CMDs the isochrones as predicted by \citetalias{H23} (purple solid lines), our single-band ANNs (red dotted lines), and our multi-band ANNs (red solid line) for UBC~1516. We also report the estimated parameters of the corresponding plotted isochrones. \citetalias{H23} and our single-band ANN predict similar parameters for the cluster. On the other side, multi-band ANN benefits from infrared 2MASS data, producing isochrones consistent across all CMDs. Note, that the difference between logAge inferred by single-band and multi-band ANNs is $\sim 1$ dex. 

\begin{figure*}
    \centering
    \includegraphics[width=\textwidth]{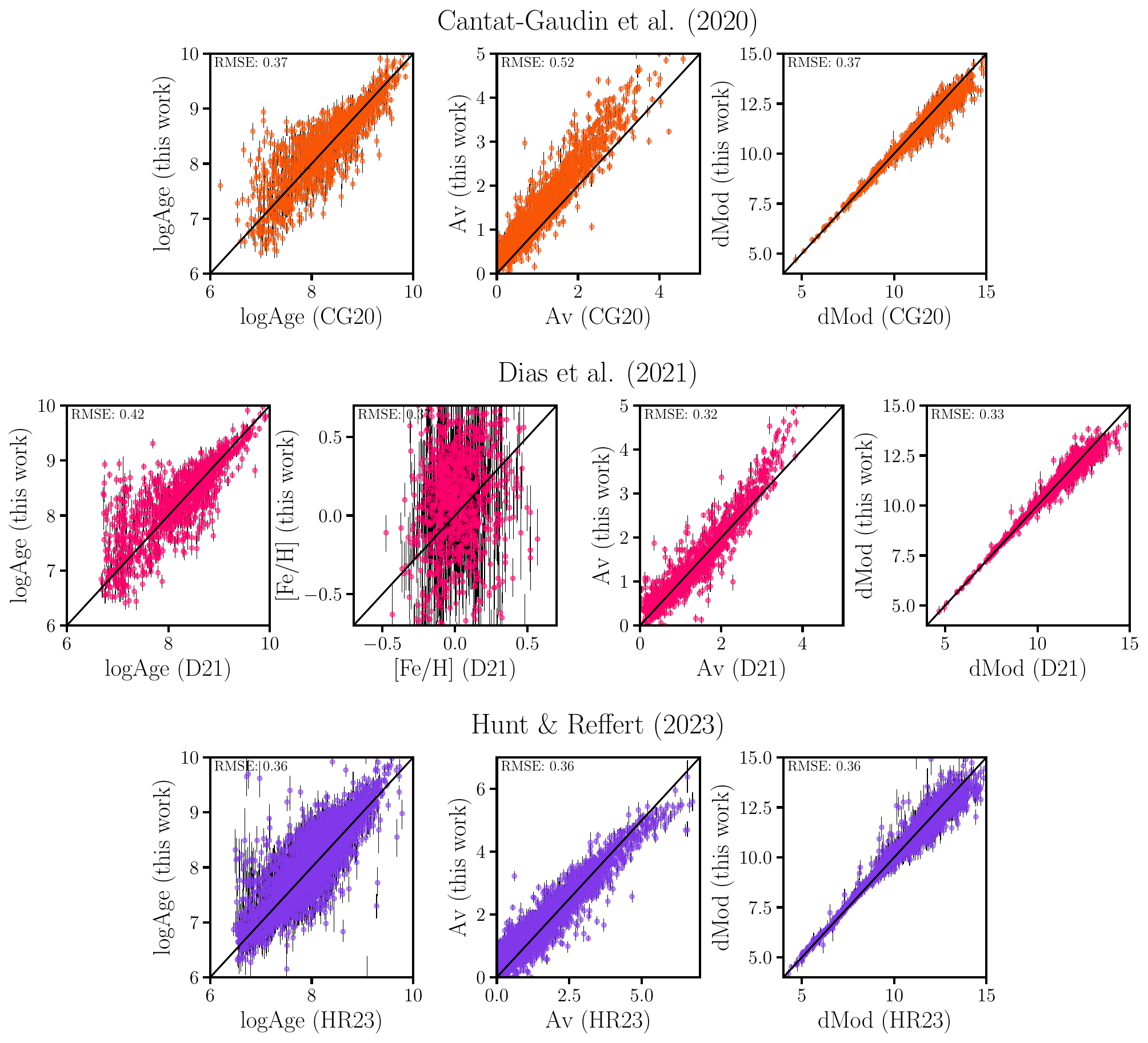}
    \caption{Comparison between parameters derived in this work with the ones present in literature. In the top left corners is reported the value of the root main square error. Top row: comparison of age, extinction, and distance for clusters that are in common with \citetalias{CG20}. Middle row: comparison of age, [Fe/H], extinction, and distance for clusters in common with \citetalias{D21}. Bottom row: comparison of age, extinction, and distance for clusters contained in \citetalias{H23}.}
    \label{fig:comparison}
\end{figure*}

\begin{figure*}
    \centering
    \includegraphics[width=\textwidth]{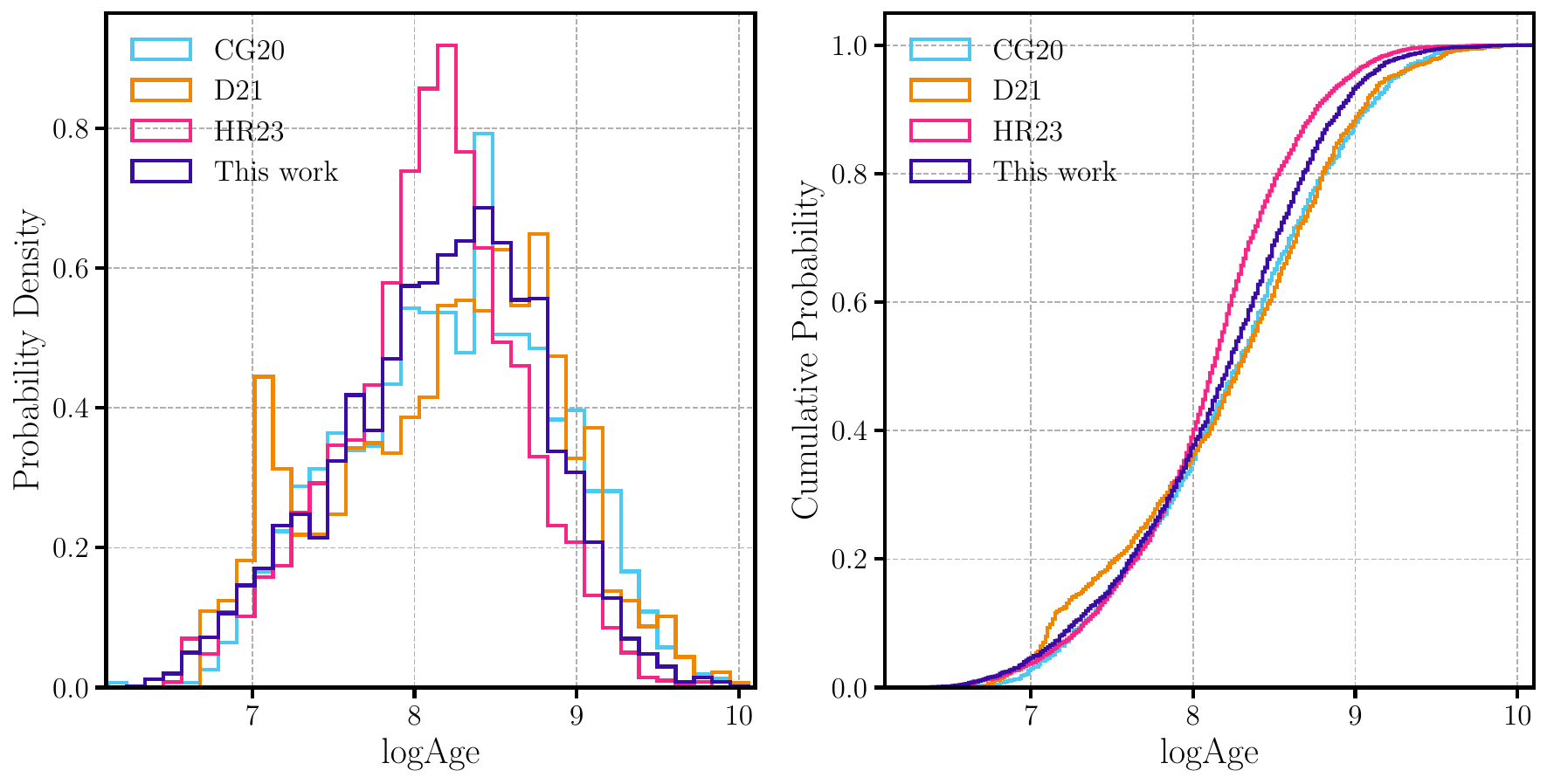}
    \caption{Age distributions of \gaia\ open clusters as predicted by \citetalias{CG20}, \citetalias{D21}, \citetalias{H23}, and in this work. Left panel: differential age distribution. Right panel: cumulative age distribution.}
    \label{fig:age_distribution}
\end{figure*}

\section{Results and comparison with the literature} \label{sec:Comparison_literature}
In this Section we compare our estimated parameters to the ones contained in three recent works present in the literature that used different approaches to the parameter estimation problem from CMDs: \citetalias{CG20}, \citetalias{D21}, and \citetalias{H23}. Our aim is to highlight the techniques used by these authors to ease the comprehension of the differences and similarities in the results. In Fig.~\ref{fig:comparison} we plot the comparison between parameters derived by \citetalias{CG20}, \citetalias{D21}, \citetalias{H23}, and our ANN for open clusters with a CMD class$>0.5$. The CMD class is the probability that a given cluster is a real one as derived by the neural network developed by \citetalias{H23}. We observe that there is a generally good agreement between our results and those from previous works. A satisfactory global agreement is also present in Fig.~\ref{fig:age_distribution}, which shows the open cluster age distribution obtained by us, and the other works from the literature. In the figure, we use open clusters with a CMD class$>0.5$ that belong to our gold sample. However, a careful inspection of Figs.~\ref{fig:comparison} and~\ref{fig:age_distribution} show interesting details and small discrepancies between these studies that here we investigate in more detail in order to get insights on the different methods of analysis.\\
In Fig.~\ref{fig:examples} we directly compare the isochrones predicted with these different methods to understand differences and their impact on the overall differences between the studies.

\subsection{Comparison with \texorpdfstring{\citetalias{CG20}}{CG20}}
\citetalias{CG20} have developed a fully connected neural network that estimates age, extinction, and distance modulus of OC using \gaia\ photometry. They have trained their neural network using a sample of real clusters whose parameters were previously determined by \citet{Bossini2019A&A...623A.108B} via isochrone fitting. Their training set was mostly composed of a sub-sample of well-known clusters whose features are then used to infer the parameters of a much larger sample of open clusters. They analysed a sample of 2017 clusters contained in previous works \citep{CGAnders2020A&A...633A..99C,Castro_Ginard2018A&A...618A..59C,Castro_Ginard2019A&A...627A..35C,Castro_Ginard2020A&A...635A..45C, LiuPang2019ApJS..245...32L} and they estimated the parameters for 1867 of them using \gaia\ DR2 photometry.\\
The top row of Fig.~\ref{fig:comparison} shows the predictions from our ANN in comparison to those obtained by \citetalias{CG20}. Focusing on the age estimates we note that some clusters being younger than 100 Myr for \citetalias{CG20} have been found to be older by our ANN. In some cases, the difference is of $\sim$1 order of magnitude in age. The cause of this discrepancy is not uniquely clear, but it is probably related to two possible pitfalls carried out by the training set that has been used.\\
As described above, the CNN developed by \citetalias{CG20} has been trained on real open clusters. This strategy has the clear advantage of training the CNN with features that are identical to those that are found in nature instead of using models and synthetic CMDs. However, since open clusters are typically younger than a few 100 Myr (see Fig.~\ref{fig:age_distribution}), their training set was not well balanced across the entire age range. That might have introduced a bias in their model favouring an age underestimation for some open clusters, in particular those where a marked turn-off is still not present (e.g., age$\lesssim$ 500 Myr; see Fig.~\ref{fig:age_distribution}). As reported in \citetalias{CG20}, the authors initially trained the ANN with synthetic CMDs. Interestingly, while the ANN was efficient at retrieving the input parameters from the synthetic CMDs, they didn't perform as well when faced with actual, observed Gaia CMDs.\\
Another possible problem is related to the method used to estimate parameters for the training set: the isochrone fitting technique \citep{Bossini2019A&A...623A.108B}. Although they focus on nearby populated clusters with very sharp CMDs. As we discuss in Section~\ref{sec:compwithD21}, isochrone fitting is also prone to age underestimation when the cluster's CMD is affected by faint contaminants.\\
The extinction values estimated by \citetalias{CG20} are systemically lower than the ones retrieved by our ANN. On average, ours are $\sim0.4$\,mag larger than the one obtained by \citetalias{CG20}. We do not find a clear explanation for this trend. However, the extinctions determined by \citetalias{CG20} are in similar disagreement also with \citetalias{D21} and \citetalias{H23}. This is caused by the fact that in the case of strong differential extinction and broadened CMDs, they tend to fit the blue edge of the sequence, ending up with lower extinctions.\\
Looking at the last plot of the top row of Fig.~\ref{fig:comparison} is seen that our ANN tends to indicate lower distances compared to \citetalias{CG20}. This can be caused by the deviation present in the extinction coefficient. The general overestimate in A$_{\rm v}$ (redder and fainter) is partially balanced by an underestimation of the distance (brighter).

\subsection{Comparison with \texorpdfstring{\citetalias{D21}}{D21}} \label{sec:compwithD21}
\citetalias{D21} used \gaia\ DR2 data to estimate age, [Fe/H], extinction, and distance modulus for 1743 open clusters retrieved via an isochrone fitting procedure using PARSEC isochrones. They analysed a sample of $2237$ clusters (for $1743$ of them they successfully determine parameters) composed of clusters previously discovered by 
\citet{CG2018A&A...618A..93C}, \citet{Castro_Ginard2019A&A...627A..35C}, \citet{LiuPang2019ApJS..245...32L}, \citet{Sim2019JKAS...52..145S}, \citet{CGAnders2020A&A...633A..99C}, \citet{Castro_Ginard2020A&A...635A..45C}, \citet{Ferreira2020MNRAS.496.2021F}, and \citet{Monteiro2020MNRAS.499.1874M}.
As priors of their analysis, they used the Galactic metallicity gradient presented in \citet{Donor2020AJ....159..199D} for [Fe/H] and the 3D extinction map developed by \citet{Capitanio2017A&A...606A..65C} for $A_{\rm V}$.\\
The comparison with the parameters derived by \citetalias{D21} is plotted in the middle row of Fig.~\ref{fig:comparison}. We report a mild disagreement between our ages and those of \citetalias{D21}. More specifically, our ANN tends to assign larger ages to some clusters that have been labelled as young (i.e., $\lesssim$100 Myr) by \citetalias{D21}. This behaviour is extremely similar to that observed in the comparison with \citetalias{CG20}. The effects are visible also in Fig.~\ref{fig:age_distribution} where both \citetalias{D21} and \citetalias{CG20} find more open clusters with ages ranging between 10 and a few tens of Myr than us. In particular, the analysis by \citetalias{D21} results in a spike at $\sim$10 Myr in the cluster age distribution. \\
\begin{figure*}
    \centering
    \includegraphics[width=\textwidth]{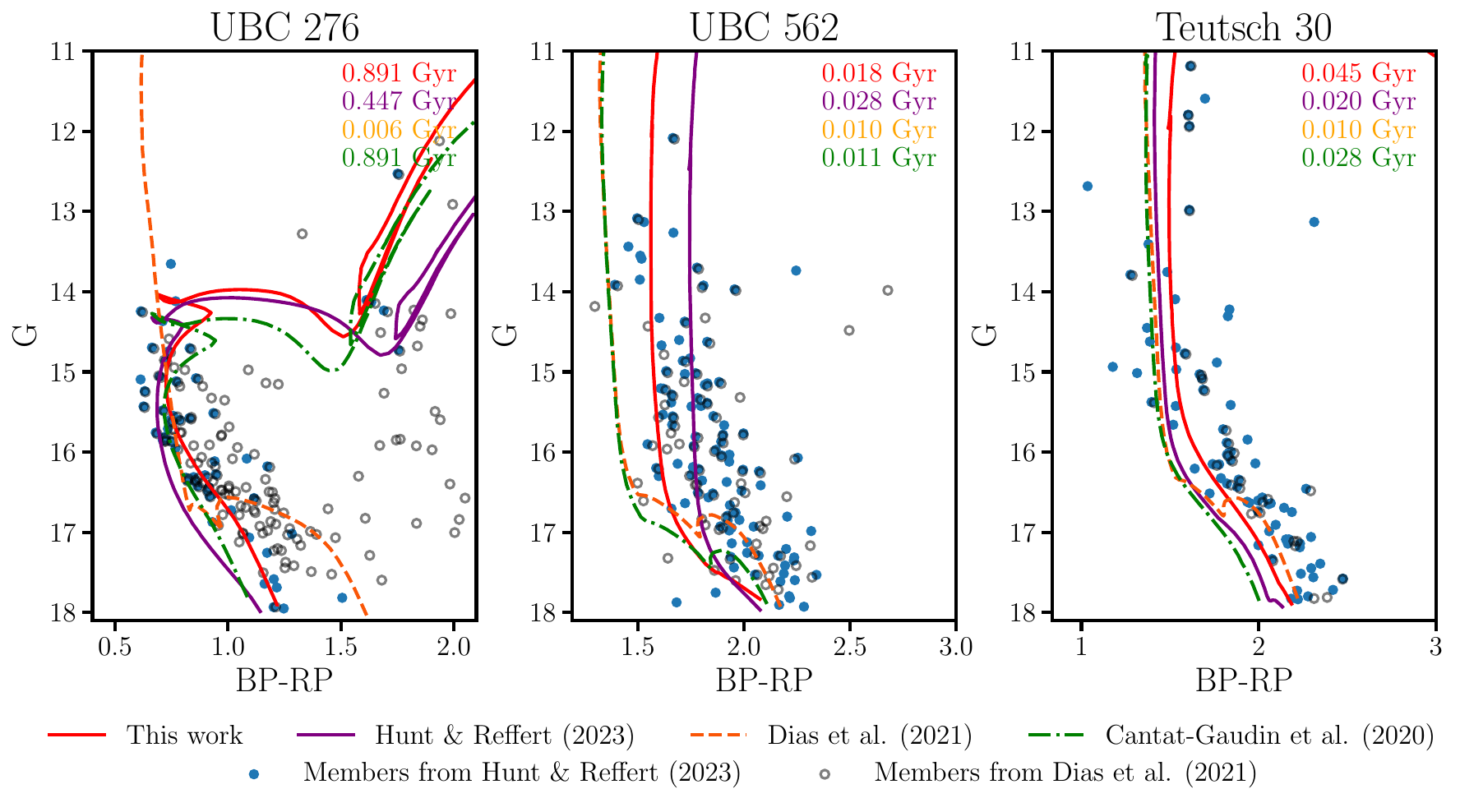}
    \caption{Predicted isochrones for three open clusters UBC 276  (left panel), UBC 562 (middle panel), and Teutsch 30 (right panel) from \citetalias{CG20} (dot-dashed green line), \citetalias{D21} (dashed orange line), \citetalias{H23} (solid purple line), and this work (solid red line). In the top right, we annotate the ages that correspond to the plotted isochrones (coloured accordingly). With blue and full dots we plot the members of the clusters retrieved by \citetalias{H23}, with empty black dots are plotted the ones of \citetalias{D21}.}
    \label{fig:examples}
\end{figure*}
During our investigation of the clusters with discrepant age determinations, we discovered that for many of these problematic cases, the membership from \citetalias{D21} (and sometimes also that from \citetalias{CG20}) contains a considerable number of faint and red contaminants which can greatly impact the performance and reliability of the isochrone fitting procedure. The isochrone fitting method is particularly susceptible to the distribution of stars in the region of the CMD where most stars are located: the low luminosity end of the sequence. That happens because the automatic fitting procedure wants to minimize the distance between the stars and the isochrone, regardless of where the star is located in the CMD. This strategy is somewhat different to an isochrone fitting carried out by eye, for which one gives more importance to very few stars at the turn-off or in the red clump instead of the multitudes in the lower main sequence. Therefore, if the cluster's sample includes a significant number of red contaminants, the automatic fitting procedure is forced to capture most of them with a prominent pre-main sequence. On the other side, the spread of faint stars at the bottom of the CMDs may be attributed not only to contamination but also to systematic effects arising from photometric errors and blends at low signal-to-noise ratios \citep{Mateo1988ApJ...331..261M, VallenariOrtolani1993ESOC...47..125V}. When working with a magnitude-limited (or S/N-limited) sample, overestimated luminosities are more likely to occur at the faint end. Moreover, due to the tilted nature of the main sequence with redder stars at fainter magnitudes, this effect can lead to an overestimation of pre-main sequence (red) stars, as discussed above.\\
An example of the bias reported above is shown in Fig.~\ref{fig:examples} (left panel), where we plot the CMD of UBC~276. Although the stellar members found by \citetalias{H23} align on a very neat sequence (blue points), those used by \citetalias{D21} for the fitting procedure (empty circles) include a large number of red contaminants. As a result, isochrone fitting by \citetalias{D21} see these latter as part of the pre-main sequence of a 6 Myr old open cluster. Instead, all the other methods result in much larger ages. Another interesting case is UBC~562 (central panel of Fig.~\ref{fig:examples}), where both the results by \citetalias{D21} and \citetalias{CG20} are off from the upper cluster sequence. However, they both succeed in fitting the lower sequence which is rich in stars, most of which are contaminants according to \citetalias{H23}. It is difficult to fully understand this strange behaviour, however, it is possible that a large contamination of non-members in the low-luminosity end of the sequence can produce deep potential wheels in the isochrone fitting loss function and forces the entire procedure to fit these low-luminosity stars with pre-main sequences.\\
Another valuable example is Teutsch 30, plotted in the right panel of Fig.~\ref{fig:examples}. In this case, \citetalias{D21} fit the low mass end of the sequence with a pre-sequence phase (widely discussed above). On the other side, both \citetalias{CG20} and \citetalias{H23} seem to follow the blue part of the CMD losing the group of bright stars at $G<12$ (members of the cluster for both \citetalias{D21} and \citetalias{H23}). We find several other cases where the isochrone fitting procedure is affected by red and faint contaminants in a very similar way. This problem - which is the cause of the spike at 10~Myr in Fig.~\ref{fig:age_distribution} - is especially important for those cases where the number of non-members is comparable to the one of "real" members. Also, only young open clusters (e.g., age$\lesssim$100~Myr) can have their ages underestimated by isochrone fitting. On the contrary, the isochrone fit method has been proven to be highly reliable for analysing old stellar populations, as it is facilitated by the presence of strong and distinct age constraints such as the turn-off, red clump, and prominent giant branches. In this sense, it is reassuring to observe a general agreement between our age estimates for older open clusters (logAge$>8.5$) and those from \citetalias{CG20} and \citetalias{D21}.\\
From the clusters plotted in Fig.~\ref{fig:examples} is possible to notice that \citetalias{H23} tends to predict isochrones slightly bluer than the ones of our ANN. This tendency will be discussed in details in Section~\ref{sec:comparisonH23}.\\
Metallicity determination solely based on CMDs is an extremely difficult task. For that reason \citetalias{D21} have fitted their isochrones imposing two strong priors on [Fe/H], and also on $A_{\rm v}$. Instead, our strategy is to let the ANN decide which is the best [Fe/H] without any bias or prior. As a result, we observe a large scatter between our [Fe/H] determinations and those of \citetalias{D21}, with a total Root Main Square Error (RMSE) that reaches $\sim$0.35 dex. However, with the values of [Fe/H] and distance modulus predicted by our ANN we are able to obtain the metallicity gradient observed in the Galaxy (see Section~\ref{sec:metallicity_gradient}).\\
Just as for [Fe/H], \citetalias{D21} have imposed priors also for the $A_{\rm v}$ parameter.  However, we notice a general agreement between us and \citetalias{D21}, with only a clear trend at high extinction factors. Namely, our ANN tends to overestimate $A_{\rm v}$ of high extinction clusters (i.e. with $A_{\rm v}>3$ mag). We do not find a clear cause of this trend.\\
The distance estimations of our ANN are in general good agreement with the ones obtained by \citetalias{D21} with the exception of a slight underestimation trend for the most distant clusters. 

\begin{figure*}
    \centering
    \includegraphics[width=0.9\textwidth]{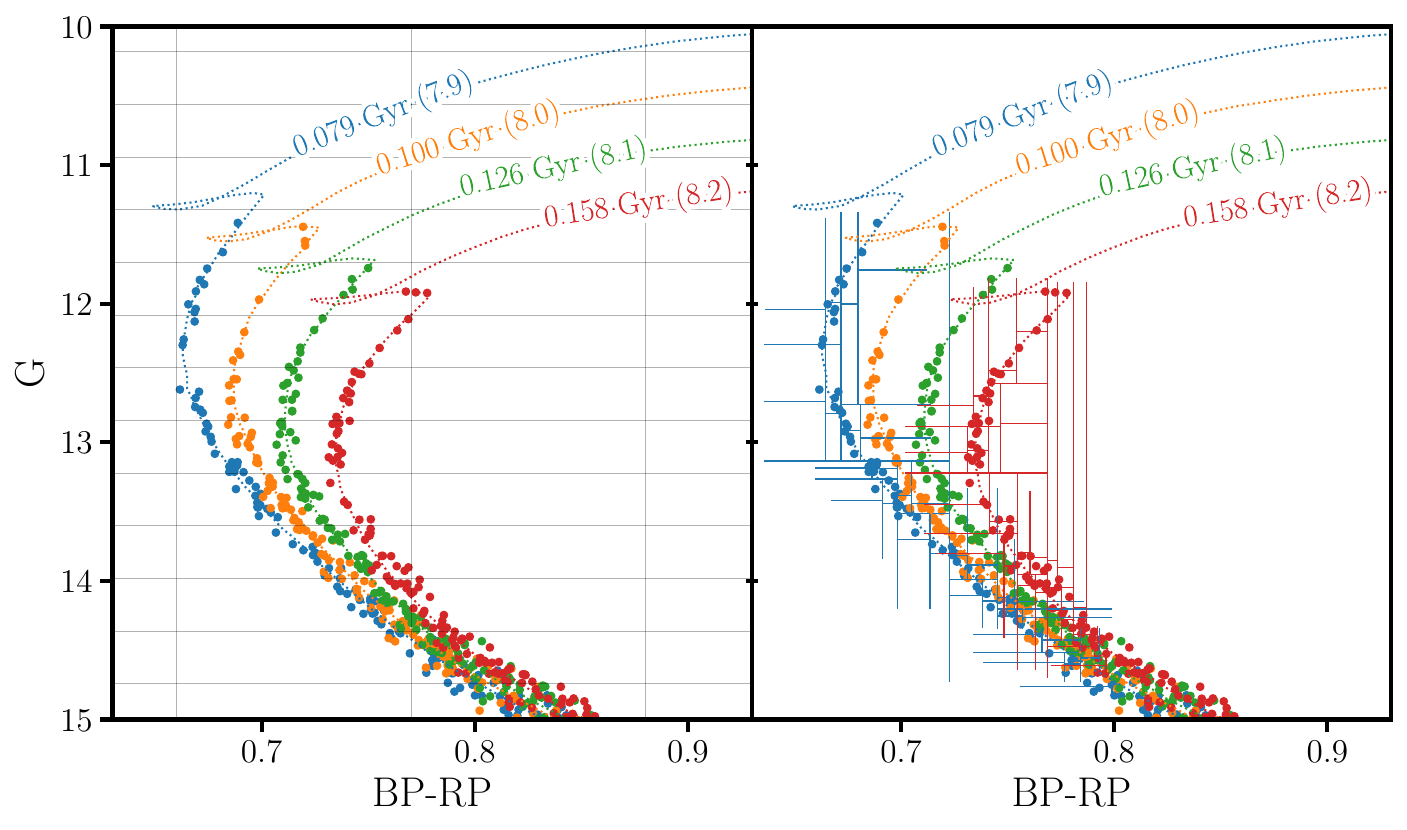}
    \caption{Synthetic CMDs generated by the code described in Section~\ref{sec:Synthetic_CMDs}. The mock cluster are generate with $N_{\rm star}=400$, [Fe/H]$=0$, $A_{\rm v}=2$, dMod$=12$ mag and logAge between $7.9$ and $8.2$. Single members of simulated clusters are plotted with dots coloured according to their age. With dotted lines, we plot the theoretical isochrones used to generate the synthetic CMDs. Left panel: synthetic CMDs superimposed to the grid used by \citetalias{H23}. Right panel: synthetic CMDs superimposed to the Quadtrees of the simulated clusters with logAge$=7.9$ (blue) and logAge$=8.2$.}
    \label{fig:gridvsquad}
\end{figure*}

\subsection{Comparison with \texorpdfstring{\citetalias{H23}}{HR23}} \label{sec:comparisonH23}
\citetalias{H23} used a quasi-Bayesian CNN trained with synthetic clusters to estimate age, extinction, and distance of clusters contained in their catalogue. In the bottom row of Fig.~\ref{fig:comparison} we only compare the parameters obtained by our ANN of cluster labelled as 'real' by \citetalias{H23} (i.e. median CMD class $>0.5$, see \citetalias{H23}). The clusters used for this comparison are the same plotted for \citetalias{H23} in the clusters' age distribution of Fig.~\ref{fig:age_distribution}.

The major discrepancy between our results and those by \citetalias{H23} is particularly evident in Fig.~\ref{fig:age_distribution} as a quasi-systematic shift in the age distributions. More specifically, the age distribution found by \citetalias{H23} is mostly shifted towards lower ages and also seems to be narrower than ours. Also, \citetalias{H23} have a lower fraction of clusters older than logAge$>8.5$ dex compared to all the other works discussed in the present paper, which are mutually consistent. This peculiarity may suggest that the CNNs of \citetalias{H23} tend to underestimate the age of clusters older than $500$ Myr considering them in the age range 8$\lesssim$ logAge $\lesssim 8.4$. Hence, their distribution peaks at ${\rm logAge}\sim8.1$, instead of ${\rm logAge}\sim8.5$ such as it is consistently found by the other works compared in Fig.~\ref{fig:age_distribution}. This difference between us and \citetalias{H23} is particularly surprising because both the studies are based on the same sample of stellar members, thus the different outputs must entirely be the products of the different approaches of analysis. Given that we both use neural networks to estimate the parameters, the difference must be due to the ways \citetalias{H23} and us extract information from the CMDs. \\
\begin{figure*}
    \centering
    \includegraphics[width=\textwidth]{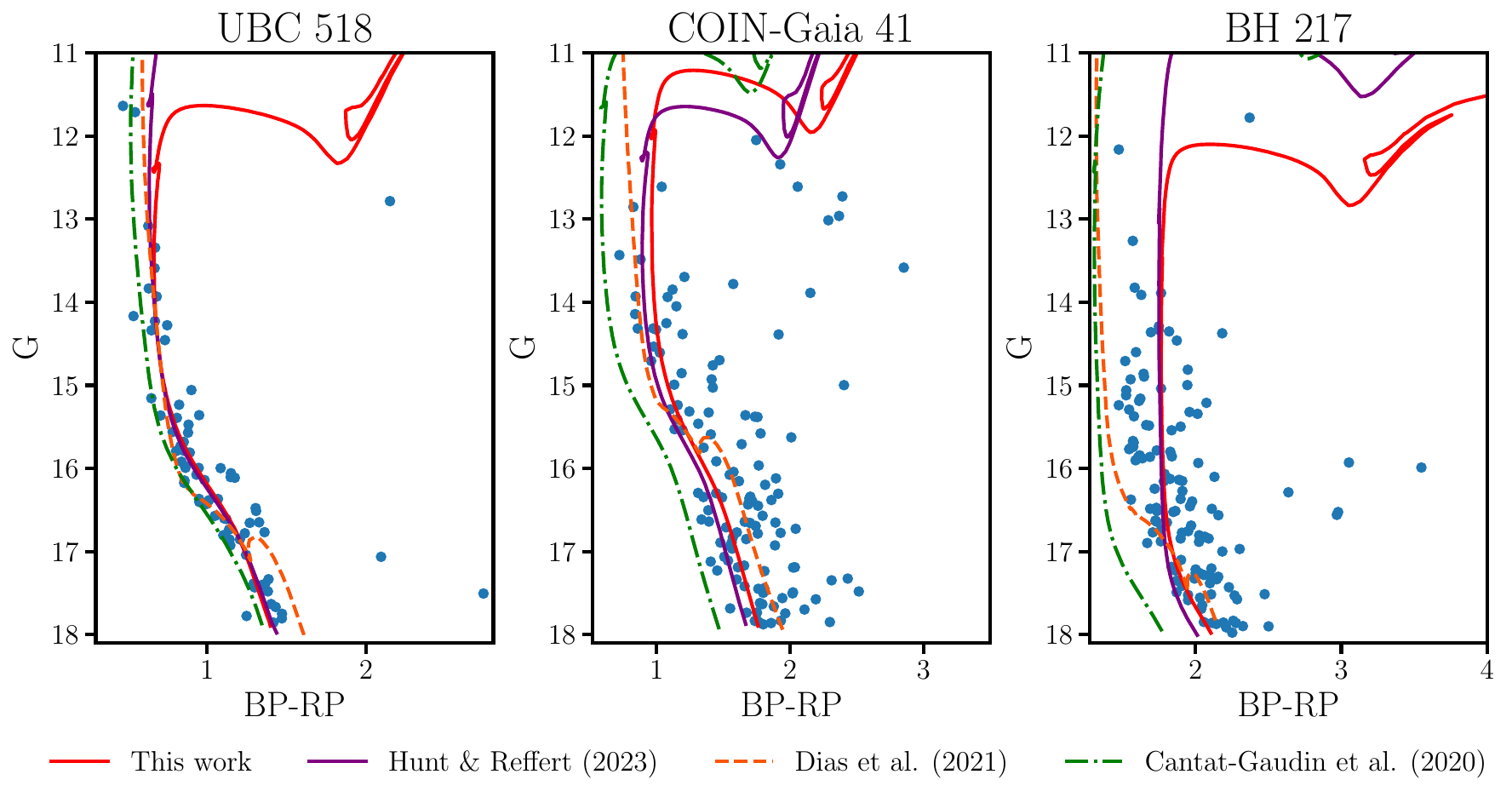}
    \caption{Predicted isochrones for three open clusters UBC 518 (left panel), COIN-Gaia 41 (middle panel), and BH 217 (right panel) from \citetalias{CG20} (dot-dashed green line), \citetalias{D21} (dashed orange line), \citetalias{H23} (solid purple line), and this work (solid red line). With blue dots, we plot the members of the clusters retrieved by \citetalias{H23}.}
    \label{fig:examples2}
\end{figure*}
More specifically, \citetalias{H23} feed their CNN with 2D histograms of the clusters' CMD. These images are formed by $32\times 32$ pixels and each pixel brings information on the density of stars falling within that small portion of the grid. The pixels have sizes of $0.38$ mag in $G$ and $0.11$ mag in ($G_{\rm BP}-G_{\rm RP}$). Instead, our strategy is to extract the relevant information using a QuadTrees algorithm (see Section~\ref{sec:QuadTrees}) whose results are then directly fed to our ANN. In Fig.~\ref{fig:gridvsquad} we show how these different methods can affect the final results. Namely, we plot four synthetic clusters with ages in the range $7.9\leq$logAge$\leq8.2$. In the left panel, we plot the pixelation grid used by \citetalias{H23}, on the right panel we show the feature extraction carried out by the QuadTrees. From the left panel of Fig.~\ref{fig:gridvsquad} is clear that all the synthetic CMDs fall into the same colour bins due to the pixelation and - even though they have different ages -  they become indistinguishable to the neural network. 

When a neural network has to indiscriminately choose between a range of outputs, it will likely choose the most probable one. Typically there are more young isochrones than old ones crossing a specific colour bin. Therefore, when the model has to choose between all the isochrones fitting a given pixel it is biased towards younger ages which are considered the most probable by the network. By contrast, the QuadTree technique produces coordinates that are always different between clusters of different ages (see right panel of Fig.~\ref{fig:gridvsquad}). In conclusion, pixelation seems not as efficient as QuadTree in exploiting at maximum the full potential of precise magnitudes produced by \textit{Gaia}. This loss of resolution is probably further magnified by the differential reddening introduced by \citetalias{H23} in their synthetic CMDs and that has the effect of making the clusters' sequences ticker. 

A consequence of this loss of resolution due to pixelation is also seen in Fig. 5 of \citetalias{H23} (top row), in which is plotted the performance of the ANN to retrieve the parameters of the test sample of clusters. In their third plot, the one on differential reddening, it is possible to notice that the model is not able to correctly recognise clusters with no extinction and therefore assigns to them a non-zero value. It is likely that their model sees the clusters with no-differential reddening and small differential reddening as indistinguishable due to the loss of resolution caused by pixelation. Therefore, it assigns at least a small amount of differential reddening to all clusters simply because that is the most $probable$ option as it has been learned by the CNN.

On top of pixelation, there is another key difference between the methods used by \citetalias{H23} and our analysis which is related to the type of neural network. More specifically, \citetalias{H23} adopt Convolutional Neural Networks (CNN) fed with images of CMDs. The convolution operation in combination with max pooling makes CNNs approximately invariant to translation. This type of inductive bias has significantly contributed to the success of CNNs for classification tasks such as image recognition, where the algorithm has to identify specific objects or patterns within an image: typically the ability to detect patterns independently from their location within the image can greatly improve their generalization ability. On the other hand, this translation invariance can hurt the performance of CNNs in cases where the position of the objects in the image matters. This is exactly the case of analysing the sequences of stars in a CMD, where positional information is critical and translation always results in different clusters' parameters. Instead, our analysis is based on Artificial Neural Networks (ANNs) fed with the arrays of numerical values generated by QuadTrees. Notably, ANNs excel in preserving most of the information passed to the input neurons. Although that can limit the generalisation ability of the algorithm, it is certainly a valuable trait for the analysis of CMDs.

Besides the systematic shifts discussed above, in the bottom row of Fig.~\ref{fig:comparison} it is possible to notice also a group of outliers, reported to be extremely young (logAge$<7$) by \citetalias{H23}, that our ANN classifies as intermediate/old age (logAge$>7.5$). These clusters are generally poorly populated and in some cases very sparse. Part of the discrepancies arise from clusters that are contained in our bronze/wood sample. In other cases, the discrepancies are not caused by the ANN approach used by us and \citetalias{H23} but rise from an intrinsic limitation of using CMDs to estimate the age of clusters. In particular, in some cases, there is the absence of clear features (e.g. sub-giant branch, turn-off, etc.) useful to remove the degeneration in age. It is clear that the limitation is exponentially higher in clusters with few members (as in these cases). As expected, part of the out-layers in logAge are also out-layers in $A_{\rm v}$ and dMod.\\

\subsection{Possible pitfalls in our analysis}
While testing our ANN we find that our model is very sensitive to the presence of stars in the red part of the CMD. This sensitivity is both a strength and a weakness, depending on the case.

We know that the red clump is a strong age indicator. Stars in the red clump have not completed the He burning and they are burning helium in the core, which is surrounded by an H-burning shell.
They have relatively similar masses and luminosities, resulting in a tight grouping in the CMD. Our ANN seems to have learned that the presence of stars in the red part of the CMD is a strong age indicator. This characteristic of our ANN is clearly visible in Fig.~\ref{fig:examples2} where we plot the isochrones inferred by \citetalias{CG20}, \citetalias{D21}, \citetalias{H23}, and our ANN for three different clusters. Focusing on UBC~518 (left panel of Fig.~\ref{fig:examples2}) our ANN retrieves an isochrone (red solid line) that is significantly different to the ones obtained by the other works. In particular, the sensibility of our ANN leads to an intermediate-age isochrone that reaches the red and bright star at G$\sim$13. If that star is a real member of the cluster we can conclude that our ANN has a better performance with respect to the other works. The cluster COIN-Gaia~41 (central panel of Fig.~\ref{fig:examples2}) is another case where the sensibility of our ANN is functional. In fact, for this cluster, our ANN is able to correctly identify the stars that belong to the red clump (if real and are not contaminants). This trait of our ANN leads to an isochrone that seems more in agreement with the distribution of stars in the CMD compared to the ones derived by \citetalias{CG20} and \citetalias{D21}. We notice that the tool developed by \citetalias{H23} has produced an isochrone that is very similar to the one obtained by us. 
\begin{figure*}
    \centering
    \includegraphics[width=0.9\textwidth]{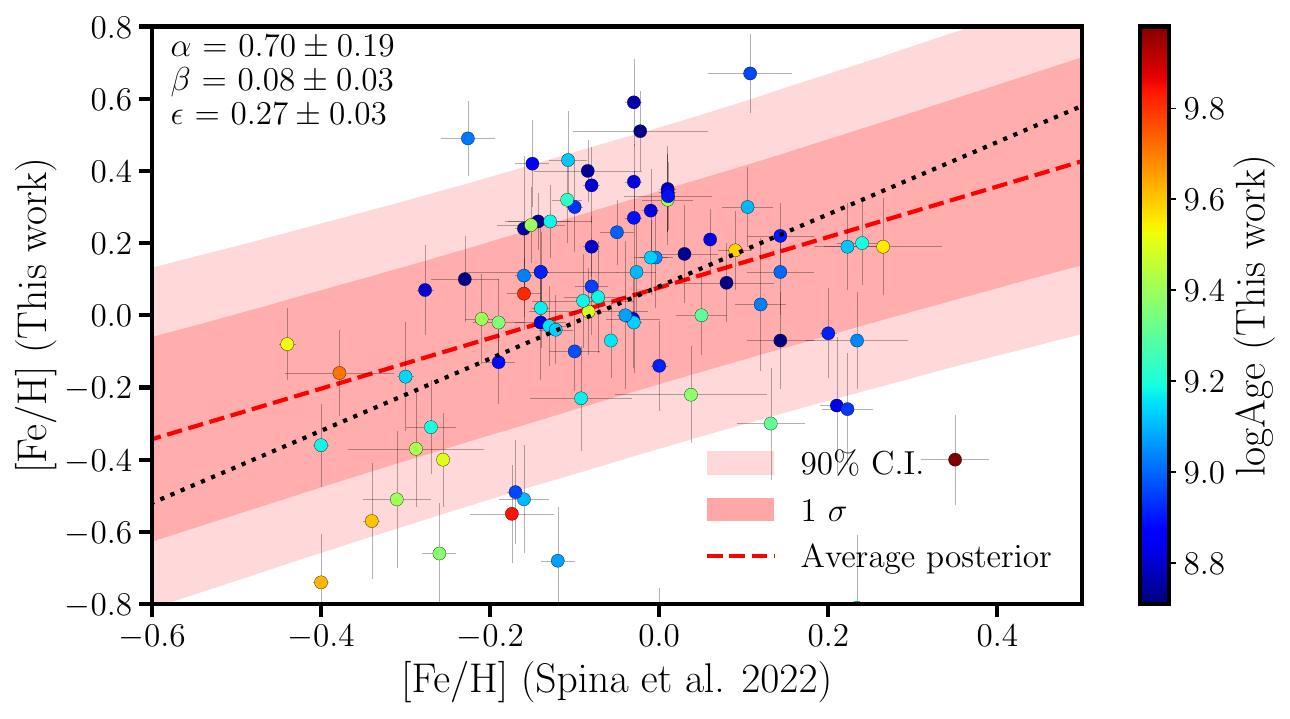}
    \caption{Comparison between [Fe/H] derived in this work and the one obtained from spectroscopic data by \citet{Spina2022Univ....8...87S}. Clusters are colour-coded based on the logAge as determined by our ANN. The average posterior of the fitted relation is plotted by a dashed red line, while the red shaded areas represent the $68\%$ and $90\%$ confidence intervals. In the top-left corner, we provide the mean and standard deviation for the three free parameters. With a dotted black line we plot the relation with $\alpha=1$ and $\beta=0.08$.}
    \label{fig:FeH_validation}
\end{figure*}

However, this high sensibility to stars on the left side of the CMDs has the drawback of being susceptible to red contaminants. That feature is visible from the CMD of BH~217 (right panel of Fig.~\ref{fig:examples2}). In that case, the isochrone inferred by our ANN is not compatible with the observed CMD. It is plausible that our ANN tends to indicate an intermediate-age isochrone with a prominent turn-off because of the presence of the group of faint and red stars (probably contaminants). It is worth noticing, that for this cluster \citetalias{H23} have obtained a more plausible isochrone (purple solid line) also compared to \citetalias{CG20}, and \citetalias{D21}.

Therefore, it is clear that the ANN developed in this work is susceptible to red contaminants in the observed CMDs. Currently, our ANN is sensible to red members almost regardless of their magnitude. Thus, even when only red and faint members are present in the CMD, the ANN tends to pass to a non-existing red clump (see right panel of Fig.~\ref{fig:examples2}). In order to mitigate this problem we use a dropout layer right after the input layer (see Section~\ref{sec:Artificial_neural_network}). However, the undesired effect is not completely removed. In future work, we plan to improve our ANN in order to avoid these misleading cases.

\section{Scientific Validation} \label{sec:Scientific_validation}
In the following, we present a comprehensive scientific validation of the proposed ANN methodology. We seek to demonstrate its effectiveness in accurately determining crucial parameters of \gaia\ OCs such as age, metallicity, extinction, and distance.\\

\subsection{Comparison with spectroscopic data}
In Section \ref{sec:compwithD21} we noticed that the [Fe/H] predictions of our ANNs have a large scatter around the one of \citetalias{D21} (derived with an isochrone fitting procedure with the galactic gradient as a prior), with a RSME $\sim0.35$ dex. To test and validate the metallicity predictions of our tool we compare them with a sample of OCs with spectroscopic data and thus [Fe/H] estimates.\\
For this test, we use the catalogue of 251 Galactic OCs produced by \citet{Spina2022Univ....8...87S}. The sample has been build through the collective analysis of homogenised data, obtained from high-resolution spectroscopic surveys and programmes, namely APOGEE \citep{APOGEE2017AJ....154...94M}, Gaia-ESO \citep{Gaia-ESO_2012Msngr.147...25G}, GALAH \citep{GALAH2018MNRAS.478.4513B}, OCCASO \citep{OCCASO_2016MNRAS.458.3150C}, and SPA \citep{SPA_2019A&A...629A.117O}. From the original sample, we select clusters with precise [Fe/H] estimates (i.e. with errors $<0.1$ dex) that are present in our analysed ones, ending up with 197 objects. We further selected those open clusters which: i) have ages greater than $500$ Myr; ii) are contained in the gold sample; iii) have at least a median CMD class$>0.5$ (i.e. reliably 'real', see \citetalias{H23}). The final test sample is composed of 89 clusters.\\
To test the correlation between the metallicities estimated by our ANNs and the one obtained from spectroscopy we use a simple linear model $y_i = \alpha \times x_i + \beta$, where $y_i$ and $x_i$ are normal distributions defined as follows: $y_i=\mathcal{N}({\rm [Fe/H]}^{\rm ANN}_i, \sigma^{\rm ANN}_{{\rm [Fe/H]},i})$ and $x_i=\mathcal{N}({\rm [Fe/H]}^{\rm S22}_i, \sigma^{\rm S22}_{{\rm [Fe/H]},i})$. $\sigma^{\rm ANN}_{{\rm [Fe/H]},i}$ and $\sigma^{\rm S22}_{{\rm [Fe/H]},i}$ are the uncertainties associated with the predictions of our ANN on the $i-$th cluster and the ones of the estimates of \citet{Spina2022Univ....8...87S}, respectively. 
Beyond $\alpha$ and $\beta$, we also introduce a third free parameter, $\epsilon$, to account for the intrinsic scatter inherent in the data. This scatter is attributed to a blend of processes that are difficult to model and/or are neglected in our modelling. For $\alpha$ and $\beta$ we chose priors that are $\mathcal{N}(1$, $1$) and $\mathcal{N}(0$ dex, $1$ dex), respectively. For the prior of the intrinsic scatter $\epsilon$ we adopt a positive half-Cauchy distribution with $\gamma=1$. Using the Python package \textsc{pymc3} \citep{Salvatier2016ascl.soft10016S} we compute 10'000 samples using a No-U-Turn Sampler \citep{Hoffman2011arXiv1111.4246H}; half of them have been discarded as part of the burn-in phase.\\
In Fig. \ref{fig:FeH_validation} we plot the selected open clusters as a function of their [Fe/H] as predicted by \citet{Spina2022Univ....8...87S} and our ANNs, with relative errors, coloured accordingly to their logAge (from our ANNs). With a red dashed line we plot the average relation, and with red shaded areas the 68$\%$ and $90\%$ confidence interval (C.I.) of the posteriors of the free parameters. In the top left corner of the figure, we report the mean values of $\alpha$, $\beta$, and $\gamma$, with relative uncertainties. From this analysis, we found a mild correlation between the [Fe/H] values, which proves that our ANN is able to extract information on the cluster metallicity from its CMDs. It is evident that the ANN does not merely produce random values within the [Fe/H] training range. Instead, it can effectively differentiate between metal-poor and solar-metallicity clusters. In particular, Fig. \ref{fig:FeH_validation} reveals a noticeable absence of points in the top-left corner, while the region around [Fe/H]$=0$ dex is densely populated.\\
Our tool is the only one currently available in the literature that is able to reliably obtain the metallicity of a cluster from its photometry and without spectroscopic data.

\subsection{Metallicity gradient} \label{sec:metallicity_gradient}
In our Galaxy, the metal content in disk stars decreases moving from the inner to the outer parts. This fact is known as the Galactic metallicity gradient. In particular, the metal content [Fe/H] decreases as the distance to the Galactic centre ($R_{\rm Gal}$) increases. In the inner part ($R_{\rm Gal}<12$ kpc) of the Galaxy the [Fe/H]-$R_{\rm Gal}$ relation is linear \citep[see][]{Spina2022Univ....8...87S}.

As a validation test, we aim to verify that the [Fe/H] and dMod parameters estimated by our ANN are consistent with the observed Galactic metallicity gradient. To do that, we first transform the distance modulus in $R_{\rm Gal}$, assuming a distance of the Sun from the Galactic Centre of $8.122$ kpc \citep{SunDistance2018A&A...615L..15G}. For this test, we selected from the sample of open clusters those which: i) have ages greater than $500$ Myr; ii) are contained in the gold sample; iii) are located from 5 to 12 kpc from the Galactic centre; iv) have at least a median CMD class$>0.5$ (i.e. reliably 'real', see \citetalias{H23}).

\begin{figure*}
    \centering
    \includegraphics[width=0.98\textwidth]{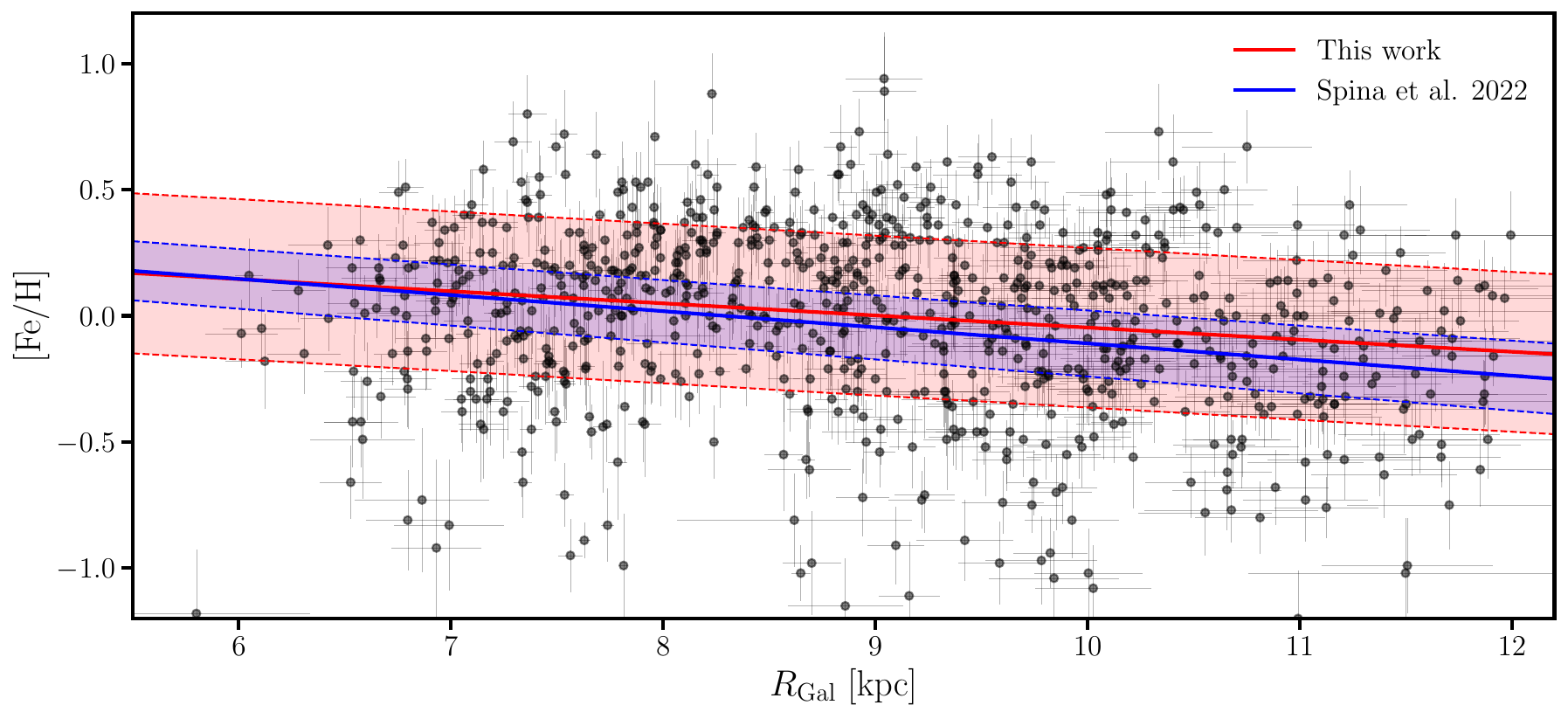}
    \caption{Values of [Fe/H] as a function of the distance from the Galactic centre ($R_{\rm Gal}$). With black dots, we plot the clusters analysed in this work and selected as reported in Section~\ref{sec:metallicity_gradient} and their respective uncertainties. With a red solid line, we plot the relation obtained by the mean values of $\alpha$ and $\beta$ while $68\%$ confidence interval is plotted with a red shadow area. We also superimpose the metallicity gradient obtained by \citet{Spina2022Univ....8...87S}, mean with a solid blue line and $68\%$ confidence interval with blue shadow area.}
    \label{fig:gradient}
\end{figure*}

We model the gradient using a simple linear model $y_i = \alpha \times x_i + \beta$, where $y_i$ and $x_i$ are normal distributions defined as follows: $y_i=\mathcal{N}({\rm [Fe/H]}_i, \sigma_{{\rm [Fe/H]},i})$ and $x_i=\mathcal{N}(R_{{\rm Gal},i}, \sigma_{R_{\rm{Gal}},i})$. $\sigma_{R_{\rm{Gal}},i}$ and $\sigma_{{\rm [Fe/H]},i}$ are the uncertainties associated with the predictions of our ANN on the $i-$th cluster. Note that in general the uncertainties on the parameters extracted by our ANN might be asymmetric. Given that the asymmetry is $\sim$ one order of magnitude less than the uncertainty itself, we compute $\sigma_{R_{\rm{Gal}},i}$ and $\sigma_{{\rm [Fe/H]},i}$ by averaging the uncertainties of our ANN. Besides $\alpha$ and $\beta$ we also fit the intrinsic scatter of data with a third free parameter $\epsilon$. The source of the scatter is a combination of processes that are difficult to model and predict such as migration, systematics, etc. For $\alpha$ and $\beta$ we chose priors that are $\mathcal{N}(-0.068{\rm \: dex}{\rm~kpc}^{-1}, 0.1{\rm \: dex}{\rm~kpc}^{-1})$ and $\mathcal{N}(0.5~{\rm dex},1~{\rm dex})$, respectively. For the prior of the intrinsic scatter $\epsilon$ we adopt a positive half-Cauchy distribution with $\gamma=1$. Using the Python package \textsc{pymc3} \citep{Salvatier2016ascl.soft10016S} we compute 10'000 samples using a No-U-Turn Sampler \citep{Hoffman2011arXiv1111.4246H}; half of them have been discarded as part of the burn-in phase.

In Table~\ref{tab:gradient_parameters} we report mean, standard deviation, and $68\%$ confidence interval (C.I.) of the posteriors of $\alpha$, $\beta$, and $\epsilon$. In Fig.~\ref{fig:gradient} we plot with a red solid line the model obtained from the mean values of $\alpha$ and $\beta$ obtained from our Bayesian analysis. The $68\%$ confidence interval is plotted with a red shadow area. We compare the gradient obtained in this work with the one retrieved from spectroscopic observations by \citet{Spina2022Univ....8...87S}. We found that the gradient estimated from the predictions of our ANN is consistent with the one obtained by \citet{Spina2022Univ....8...87S}. We also found that in our case the intrinsic scatter $\epsilon$ is $\sim3$ times larger than the one found by \citet{Spina2022Univ....8...87S}. 
The discrepancy likely arises from our estimation of metallicity using the CMD based on photometric data, which is less precise than determinations made through spectroscopy. Additionally, the uncertainties in the parameters might be underestimated, further contributing to the difference. Nevertheless, this result demonstrates that the advancement of our method (i.e. including [Fe/H] as an output parameter) is real. The information on metallicity can be reliably extracted by the sequence of stars in the CMD built on \gaia\ and 2MASS photometry.

\begin{figure*}
    \centering
    \includegraphics[width=\textwidth]{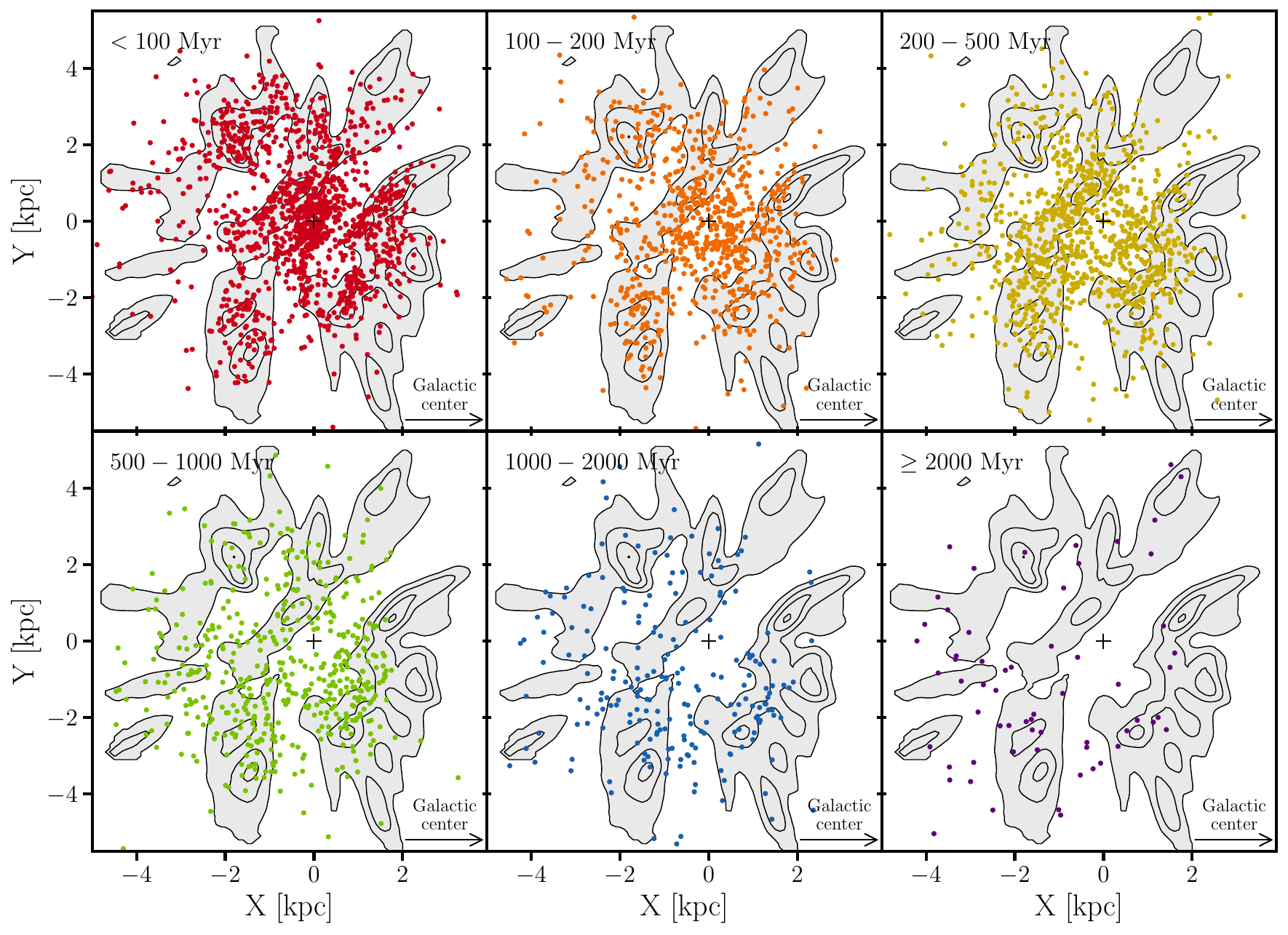}
    \caption{Spatial distribution of intrinsically bright open clusters divided in age bins superimposed to the over-density map (in grey) measured by \citet{Poggio21}. The Sun is located in $({\rm X},{\rm Y})=(0,0)$, the Galactic Center is to the right, and Y is taken as positive in the direction of Galactic rotation}
    \label{fig:oc_distribution}
\end{figure*}
\begin{figure}
    \centering
    \includegraphics[width=0.5\textwidth]{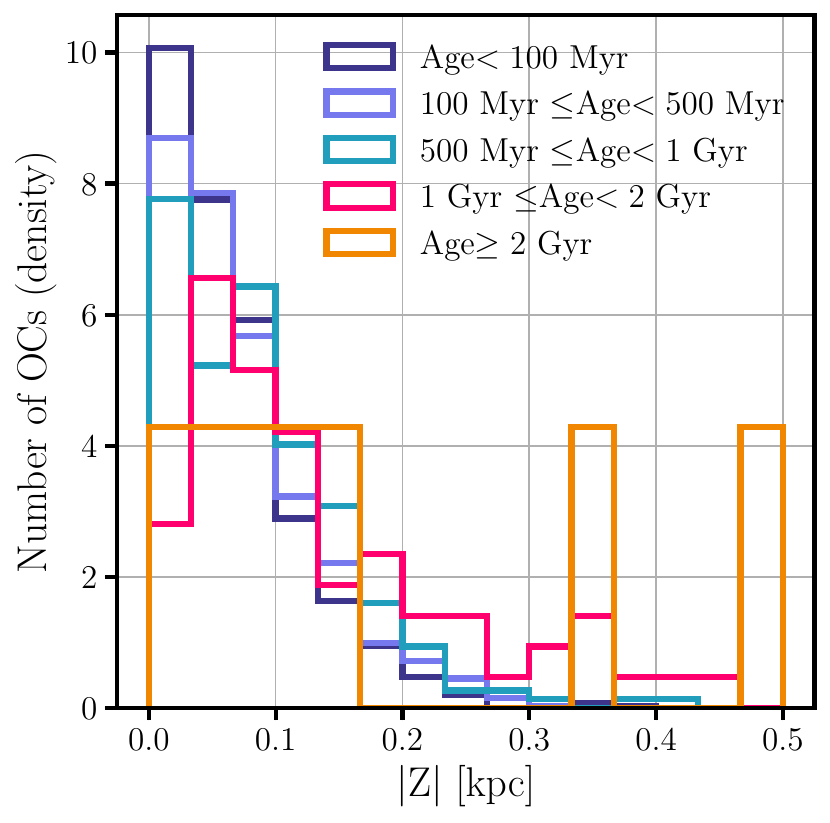}
    \caption{Distribution of heights above the Galactic midplane for intrinsically bright open clusters in different age bins.}
    \label{fig:height_scale}
\end{figure}

\subsection{The distribution of open clusters within the Galactic disk } \label{sec:disk}
The spatial distribution of open clusters within the Galactic disk is an important feature that we can inspect in order to validate our results. In Fig.~\ref{fig:oc_distribution} we plot the distribution of open clusters as a function of the Galactic cartesian coordinates X and Y for six different age bins. Among the entire sample, here we only show the open clusters that i) are included in the gold sample; ii) have a CMDclass$>$0.5; iii) have at least 3 stellar members with absolute magnitude lower than 3 dex. This latter condition allows us to exclusively select the intrinsically bright open clusters, which is necessary to highlight the spatial features formed by open clusters in the disk \citep[see, ][]{Poggio21, CantatGaudin22,Drimmel2023A&A...670A..10D, Gaia_Collaboration2023A&A...674A..37G}. In fact, a significant number of known stellar clusters are small and young associations located very close to the Sun. Some of them remain visible in the upper-left panel even after the cut in magnitude. In Fig.~\ref{fig:oc_distribution} we also show the position of the spiral arms from upper main sequence stars in \citet[][]{Poggio21}, shown as grey-shaded areas and contours. Starting from the upper left of these panels the contours show the Perseus Arm, the Local Arm and finally the Sag-Car Arm in the lower right corner. From the comparison between the location of open clusters and spiral arms, we notice that younger clusters are more concentrated near the spiral arms. It is particularly evident the clump of clusters younger than 100 Myr located in the Perseus Arm at (X,Y)$\sim$$(-1.7,2)$. Also, the higher density of clusters in the Local and Sag-Car arms is visible for ages up to 500 Myr. That is an expected behaviour as spiral arms enhance the formation of giant molecular clouds, hence stellar associations. Since young clusters had no time to migrate and disperse throughout the Galactic disk, they are still close to the spirals: the large-scale structures from which they were born. Instead, as we move towards the older clusters, these latter are more randomly distributed across the Galactic disk. It is therefore plausible that old clusters have significantly migrated from their original birthplaces.
\begin{table}
	\centering
	\caption{Mean, $\sigma$, and $68\%$ confidence interval of $\alpha$, $\beta$, and $\epsilon$ obtained from our Bayesian analysis.}
	\label{tab:gradient_parameters}
	\begin{tabular}{lccc} 
		\hline
		\textbf{Parameter} & \textbf{Mean} & \textbf{$\sigma$} & \textbf{$68\%$ C.I.}\\
        \hline
        $\alpha$ [kpc$^{-1}$]& -0.048 & 0.009 & [-0.057 , -0.039]\\
		$\beta$ & 0.43 & 0.08 & [0.35, 0.51]\\
		$\epsilon$ & 0.313 & 0.009 & [0.303, 0.322]\\
		\hline
	\end{tabular}
\end{table}
In Fig.~\ref{fig:height_scale} we show the distributions of heights above the plane for open clusters in different age bins. Here we consider the same open clusters plotted in the left panel and with a distance from the Sun that is less than 2 kpc. Here we notice that the youngest clusters are spatially concentrated on the Galactic midplane. However, as we move to older clusters the distribution spreads at larger heights. The distribution of the oldest age bin is nearly flat. Similarly to what is observed in Fig.~\ref{fig:oc_distribution} from the comparison with the spiral arms, this general behaviour described above is widely expected. Open clusters typically form very close to the midplane where larger gas reservoirs are available for star formation. Once an open cluster is formed and gets older, it has time to migrate far above the Galactic midplane \citep[e.g.][]{Migration_1_2021A&A...647A..19T, Migration_2_2012A&A...541A..64J, Viscasillas2023arXiv230917153V}. On top of that, the open clusters living to close the disk have higher chances of being disrupted by perturbations of the Galactic potential, such as giant molecular clouds and spiral arms. All of that contributes to widening the distributions of older clusters.

\section{Conclusions} \label{sec:Conclusions}
In this work, we perform the parameter estimation of the stellar cluster ($\sim 6400$) contained in the sample presented by \citetalias{H23} using colour-magnitude diagrams from both \gaia\ and 2MASS. To do that, we use a group of artificial neural networks trained on synthetic clusters based on PARSEC isochrones. To extract the features for our neural networks, we pre-process the colour-magnitude diagrams using the QuadTree algorithm. Our conclusions can be summarised as follows:
\begin{itemize}
    \item We obtain credible estimates of age, metallicity, extinction, and distance for $\sim5400$ clusters (i.e. the ones included in our gold sample). These parameters pass through a comprehensive scientific validation that demonstrates the reliability of our artificial neural network to determine the vital parameters of \gaia\ open clusters. We compared the metallicity estimates of our ANNs with ones obtained from spectroscopic studies obtaining promising results. Furthermore, with our parameters, we are able to reconstruct the Galactic metallicity gradient visible in the inner part of the disk. These results demonstrate that our method is able to reliably extract the information on metallicity from the colour-magnitude diagram built on \gaia\ photometry. With the parameters obtained from our network, we derive the 3D distribution of clusters around the Sun. From that we found that young clusters are the closest tracers of the spiral arms of the Galactic disk \citep[see][]{Poggio21} and that there is a clear increase in the scale height of clusters with their age. Both of these things are widely expected by current models and thus they represent a significant scientific validation.
    \item We compare our results with the parameters obtained by previous works that used a wide variety of techniques: \citetalias{CG20}, \citetalias{D21}, and \citetalias{H23}. We thus investigate the possible sources of some discrepancies between this work and the ones cited above. The isochrone fitting technique used by \citetalias{D21} suffers from faint and red star contaminants and in these cases tends to fit a pre-sequence that is not present in the observed CMD. \citetalias{CG20} ANN has been trained with observed CMDs with known parameters (obtained via isochrone fitting) thus the trained neural network is potentially biased as the training sample. The use of a grid in a CMD to extract its features to feed an ANN, a procedure adopted by both \citetalias{CG20} and \citetalias{H23}, seems to lose some important characteristics of the sequences.\\
    With our analysis, we find systematically older age compared to the previous works presented here. This is an interesting result, with some possible relevant specific cases.
    
    \item In this work, we make use of a novel approach to the feature extraction task by adopting a QuadTree algorithm. The QuadTree is capable of efficiently tracing the sequence even with photometric errors and outliers. The use of a QuadTree algorithm to extract numerical features from the CMDs paths the way to a second key element of novelty in the analysis, which is the use of ANNs. Previous works based on machine learning techniques have made use of CNNs (see, \citetalias{CG20} and \citetalias{H23}). However, these types of algorithms are approximately invariant to translation. That specific trait of CNNs is due to convolution and max pooling operations upon which these algorithms are designed, but also to the pixelation of the input CMDs images which necessarily results in a loss of spatial resolution. The translation invariance can hurt the analysis of sequences of stars in CMDs. Instead, ANNs fed with numerical arrays preserve the full positional information which is expected to improve the performances. Finally, with our analysis, we also demonstrate that clusters' parameters can be derived from the simultaneous analysis of multiple photometric bands. This prospect will be particularly useful in view of next-generation telescopes, such as the Vera Rubin Observatory which will provide photometry from six filters for $\sim$20$\times$10$^9$ stars.\\
    \item In the last years clustering algorithms have found more and more new clusters so it become crucial to develop parameters and/or techniques to discriminate if a group of stars is a real cluster. We found that finding an isochrone that fits the CMD of a generic cluster is a necessary but not sufficient condition to classify a cluster as real. Further analysis and studies need to be done in the future to address this crucial problem.\\
\end{itemize}
In recent years, a series of works \citetalias{CG20}, \citetalias{H23}, and this work have proved that artificial neural networks are powerful and reliable tools to analyse \gaia\ open clusters. They also proved performances similar (superior?) to the isochrone fitting technique while having lower computational costs. On the other hand, isochrone fitting methods are costly in terms of computational time because of the large parameter space that they need to explore. For this reason, it is common to assume strong priors on (some) parameters \citepalias[e.g. see][]{D21} that can introduce some biases. One possibility is to merge the two techniques taking advantage of the strong points of each one. In particular, is possible to perform a preliminary analysis by means of an ANN-based tool (e.g. similar to the one presented in this work) that can generate the distribution of the plausible parameters. The estimates of the ANN are then refined by taking them as priors of an isochrone fit procedure. This new approach will be developed and tested in future works. Future improvements to the model outlined in this work will incorporate a more sophisticated and precise modelling of synthetic CMDs, taking into account factors like binaries, stellar rotation, and differential extinction.\\

\section*{Acknowledgments}
We warmly thank the anonymous referee for their constructive and beneficial feedback, which has notably enhanced our manuscript.\\
L. C. thanks Dr. Laura Greggio for the insightful discussion on stellar evolution and stellar cluster analysis.\\
This research has made use of the \href{http://svo2.cab.inta-csic.es/theory/fps/}{SVO Filter Profile Service} supported from the Spanish MINECO through grant AYA2017-84089.\\
M. P. acknowledges financial support from the European Union’s Horizon 2020 research and innovation program under the Marie Skłodowska-Curie grant agreement No. 896248. SL acknowledges financial support by the PRIN INAF 2019 grant ObFu 1.05.01.85.14 (“Building up the halo: chemo-dynamical tagging in the age of large surveys”). J.S-U was supported by the National Agency for Research and Development (ANID)/Programa de Becas de Doctorado en el extranjero/DOCTORADO BECASCHILE/2019-72200126.

%

\vspace{5mm}







\bibliography{sample631}{}
\bibliographystyle{aasjournal}



\appendix
\section{Our catalogue of open clusters parameters} \label{sec:appendix}
In the following are listed and described all the entrances of the table available online (via \url{https://phisicslollo0.github.io/cavallo23.html}) that contains the catalogue of 6413 clusters analysed in this work.
\begin{table}[h!]
    \centering
    \caption{Description of the columns in the table available online}
    \label{table:fancy_table}
    \begin{tabular}{cccl}
        \toprule
        \textbf{Column} & \textbf{Label} & \textbf{Unit} & \textbf{Description}\\
        \midrule
        1 & Cluster                   &   -   & Cluster name\\
        2 & kind                      &   -   & [omg] Type of object (o: open cluster, m: moving group, g: globular cluster) from \citetalias{H23}\\
        3 & RA                        & [deg] & Mean right ascension of cluster's members\\
        4 & DEC                       & [deg] & Mean declination of cluster's memebers\\
        5 & plx                       & [mas] & Mean parallax of cluster's members\\
        6 & e$\textunderscore$plx     & [mas] & Error on the parallax\\
        7 & CMDclass                  &   -   & [0/1] 50th percentile of CMD class from \citetalias{H23}\\
        8 & quality                   &   -   & [0/3] Quality of the predictions (0: gold sample 1: silver sample 2:bronze sample 3:wood sample)\\
        \midrule
        9 & logAge$\textunderscore$16 & [dex] & 16th percentile of logAge predictions of the ANNs\\
        10 & logAge$\textunderscore$50 & [dex] & 50th percentile of logAge predictions of the ANNs\\
        11& logAge$\textunderscore$84 & [dex] & 84th percentile of logAge predictions of the ANNs\\
        12& FeH$\textunderscore$16    & [dex] & 16th percentile of [Fe/H] predictions of the ANNs\\
        13& FeH$\textunderscore$50    & [dex] & 50th percentile of [Fe/H] predictions of the ANNs\\
        14& FeH$\textunderscore$84    & [dex] & 84th percentile of [Fe/H] predictions of the ANNs\\
        15& Av$\textunderscore$16     & [mag] & 16th percentile of V-band extinction predictions of the ANNs\\ 
        16& Av$\textunderscore$50     & [mag] & 50th percentile of V-band extinction predictions of the ANNs\\
        17& Av$\textunderscore$84     & [mag] & 84th percentile of V-band extinction predictions of the ANNs\\
        18& dMod$\textunderscore$16   & [mag] & 16th percentile of distance module predictions of the ANNs\\
        19& dMod$\textunderscore$50   & [mag] & 50th percentile of distance module predictions of the ANNs\\
        20& dMod$\textunderscore$84   & [mag] & 84th percentile of distance module predictions of the ANNs\\
        \midrule
        21& X                         & [kpc] & X coordinate in galactocentric galactic coordinates (as predicted by the ANNs)\\
        22& Y                         & [kpc] & Y coordinate in galactocentric galactic coordinates (as predicted by the ANNs)\\
        23& Z                         & [kpc] & Z coordinate in galactocentric galactic coordinates (as predicted by the ANNs)\\
        24& Rgc                       & [kpc] & Distance from Galactic centre (as predicted by the ANNs and assuming Sun Rgc = 8.122 kpc)\\
        \bottomrule

    \end{tabular}
\end{table}

\end{document}